\documentclass[a4paper,fleqn]{cas-dc}

\ExplSyntaxOn
\cs_gset:Npn \__first_footerline:
  { \group_begin: \small \sffamily  \group_end: }
\ExplSyntaxOff 

\usepackage{amssymb}
\usepackage{framed}
\usepackage{hyperref}
\usepackage{graphicx}
\usepackage{url}
\usepackage{xcolor}
\usepackage{soul}
\usepackage[ruled,vlined]{algorithm2e} 
\usepackage{commath} 
\usepackage{tabularx} 
\usepackage{makecell} 
\usepackage{mathtools} 
\usepackage{multirow} 
\usepackage{numprint} 
\usepackage{multicol} 

\usepackage{pifont}

\usepackage[square,sort,comma,numbers]{natbib}

\usepackage{eurosym,comment}
\usepackage{multirow}
\usepackage{threeparttable}
\usepackage{verbatim}
\usepackage{subcaption}
\usepackage{float}
\usepackage{microtype}
\usepackage{colortbl}
\usepackage{amsmath}
\usepackage[british]{babel}
\setcitestyle{square}

\usepackage{array} 

\graphicspath{{Images/}}

\DeclareGraphicsExtensions{.pdf}

\newcolumntype{?}{!{\vrule width 2.5\arrayrulewidth}}
\newcolumntype{^}{!{\vrule width 1.8\arrayrulewidth}}
\newcommand{\thline}{\Xhline{2.5\arrayrulewidth}}
\newcommand{\mhline}{\Xhline{1.8\arrayrulewidth}}

\newcommand*{\ie}{i.e.}
\newcommand{\cellgps}{2.5pt}
\newlength{\Oldarrayrulewidth}

\usepackage{booktabs}
\usepackage{etoolbox}
\usepackage{soul} 

\newtoggle{finalPaper}

\setstcolor{red}

\toggletrue{finalPaper} 

\iftoggle{finalPaper} {
	\newcommand{\addtxtRev}[1]{#1}
	\newcommand{\changeRev}[2]{#2}
	\newcommand{\rmvtxtRev}[1]{}
	\newcommand{\addtxt}[1]{#1}
	\newcommand{\change}[2]{#2}
	\newcommand{\rmvtxt}[1]{}
	}{
	\newcommand{\addtxtRev}[1]{\textcolor{red}{#1}}
	\newcommand{\changeRev}[2]{\st{#1}\textcolor{red}{#2}}
	\newcommand{\rmvtxtRev}[1]{\st{#1}}
	\newcommand{\addtxt}[1]{#1}
	\newcommand{\change}[2]{#2}
	\newcommand{\rmvtxt}[1]{}
}

\begin{document}
\let\WriteBookmarks\relax
\def\floatpagepagefraction{1}
\def\textpagefraction{.001}
\shorttitle{Federated Learning for Malware Detection in IoT Devices}
\shortauthors{Rey et~al.}
\title[mode = title]{Federated Learning for Malware Detection in IoT Devices}

\author[1]{Valerian Rey}[orcid=0000-0003-3078-3598]

\author[2]{Pedro Miguel {S\'anchez S\'anchez}}[orcid=0000-0002-6444-2102]
\cormark[1]

\author[3]{Alberto {Huertas Celdr\'an}}[orcid=0000-0001-7125-1710]

\author[4]{G\'er\^ome Bovet}

\author[1]{Martin Jaggi}[orcid=0000-0003-1579-5558]

\address[1]{
École Polytechnique Fédérale de Lausanne (EPFL), 1015 Lausanne, Switzerland}

\address[2]{Department of Information and Communications Engineering, University of Murcia, Murcia 30100, Spain}

\address[3]{Communication Systems Group (CSG), Department of Informatics (IfI), University of Zurich UZH, 8050 Zürich, Switzerland}

\address[4]{Cyber-Defence Campus, armasuisse Science \& Technology, 3602 Thun, Switzerland}

\cortext[cor1]{Corresponding author.
Email address: pedromiguel.sanchez@um.es (P.M.S. S\'anchez)}


\maketitle

\begin{abstract}
Billions of IoT devices lacking proper security mechanisms have been manufactured and deployed for the last years\addtxt{, and more will come with the development of Beyond 5G technologies}. Their vulnerability to malware has motivated the need for efficient techniques to detect infected IoT devices inside networks. \change{With privacy becoming a major concern in recent years, a new technology called federated learning emerged}{With data privacy and integrity becoming a major concern in recent years, \addtxtRev{increasing with the arrival of 5G and Beyond networks,} new technologies such as federated learning and blockchain emerged}. \change{It}{They} allow\rmvtxt{s} training machine learning models with decentralized data while preserving its privacy by design. This work investigates the possibilities enabled by federated learning concerning IoT malware detection and studies security issues inherent to this new learning paradigm. In this context, a framework that uses federated learning to detect malware affecting IoT devices is presented. N-BaIoT, a dataset modeling network traffic of several real IoT devices while affected by malware, has been used to evaluate the proposed framework. Both supervised and unsupervised federated models (multi-layer perceptron and autoencoder) able to detect malware affecting seen and unseen IoT devices of N-BaIoT have been trained and evaluated. Furthermore, their performance has been compared to two traditional approaches. The first one lets each participant locally train a model using only its own data, while the second consists of making the participants share their data with a central entity in charge of training a global model. This comparison has shown that the use of more diverse and large data, as done in the federated and centralized methods, has a considerable positive impact on the model performance. Besides, the federated models, while preserving the participant's privacy, show similar results as the centralized ones. As an additional contribution and to measure the robustness of the federated approach, an adversarial setup with several malicious participants poisoning the federated model has been considered. The baseline model aggregation averaging step used in most federated learning algorithms appears highly vulnerable to different attacks, even with a single adversary. The performance of other model aggregation functions acting as countermeasures is thus evaluated under the same attack scenarios. These functions provide a significant improvement against malicious participants, but more efforts are still needed to make federated approaches robust.

\end{abstract}

\section{Introduction}
\label{sec:introduction}


By 2025, forecasts estimate that there will be about 64 billion IoT devices online \cite{riad2020dynamic}. The massive deployment of these devices is undoubtedly transforming the world into a hyper interconnected environment. The IoT paradigm, together with new \addtxt{5G and Beyond 5G (B5G)} network technologies\rmvtxt{such as 5G}, are enabling new application scenarios and businesses not seen before, such as Industries 4.0 and Smart Cities, among many others \addtxt{\cite{stergiou2020iot}}. However, simultaneously to the advances in new technologies, the number and variety of cyberattacks have grown in recent years, making current security approaches outdated in a short time \addtxt{\cite{adat2018security}}. For this reason, controlling the security of future network environments \addtxt{enabled by B5G technologies} presents open challenges that must be solved with modern techniques.


One strategy that has gained relevance when detecting devices that have been corrupted by malware is monitoring device activities to generate behavioral fingerprints or profiles. Fingerprints can be utilized to detect deviations caused by cyberattacks or malicious software modifications \cite{device_behavior}. In IoT devices, heterogeneous behavior sources can be monitored, such as network communications, resource consumption, software actions and events, or users' interactions. Therefore, depending on the objective to be achieved, one or another can be used. Concretely, when it comes to detecting cyberattacks, the most widely used dimension in the literature is network communications \cite{iot_intrusion_survey}.

Once behavior sources are selected and monitored, the next step \rmvtxt{in order }to achieve \rmvtxt{a }successful malware detection is to process the data and generate device behavior fingerprints. In \changeRev{this}{the 5G and B5G} context, Artificial Intelligence (AI) techniques, mainly Machine Learning (ML) and Deep Learning (DL), have gained enormous relevance in recent years \cite{qu2020empowering}. Nowadays, most of the existing solutions that use ML/DL to detect malware rely on a central entity in charge of collecting data from different devices and training global models. Later, these models are distributed between individual clients, or these clients send their live test data to the server for behavior evaluation and malware detection. However, this approach is not suitable for scenarios where device behaviors contain sensitive or confidential data that would significantly affect environmental security and privacy in case of falling into malicious hands. A similar situation occurs in scenarios where the monitored data sources are related to human beings and private actions are involved.


In such a context where data privacy \addtxt{and integrity are} critical, Federated Learning (FL) \cite{fedavg} \change{is}{and Blockchain are} gaining huge relevance in the last years as a collaborative ML paradigm.
In FL, the algorithm training is performed in a decentralized manner by different nodes, or clients, that use local data. In this scenario, each decentralized node trains an individual model using its own data and shares the model parameters (instead of the data) with the rest. The exchange of the model parameters and their aggregation to create a unique and global model can be performed through a central entity, called server, or following a peer-to-peer approach \cite{fl_concepts}. After several iterations, each client has a global model obtained as an aggregation of the individual model of each client. This approach enables data privacy by design, as data is not shared with any external identity.


Despite the novelty and benefits of FL approaches, their application in real-world scenarios still presents several open questions that must be analyzed and solved (or at least improved) \addtxtRev{\cite{liu2020federated}}. \changeRev{It is the case of applying FL to an equally modern field, such as the behavior analysis of IoT devices in the cybersecurity context.}{Previous works dealing with FL for intrusion detection \cite{fed_iiot,fl_iiot,blockchain_fl_id} lack the use of realistic datasets in the FL context, the analysis on adversarial impact, or the discussion of their deployment in B5G scenarios, among others.} In this sense, some of the most relevant open challenges can be summarized as: 1) how can FL be used in the IoT context to build joint models without sharing sensitive data?; 2) how do FL approaches affect the performance of traditional anomaly detectors and classifiers in IoT scenarios?; 3) what is the impact of different adversarial attacks affecting federated models designed to detect cyberattacks on IoT scenarios?; and 4) are existing countermeasure mechanisms able to mitigate the effects of adversarial attacks?; and if so, 5) what are the most suitable countermeasures for IoT scenarios?; \addtxtRev{6) how these solutions could be incorporated in future networks such as B5G?}.


With the goal of overcoming the previous open challenges, this paper presents the following main contributions:

\begin{itemize}
    \item \addtxtRev{A use case presenting a B5G scenario where there is a necessity of detecting cyberattacks affecting IoT devices, managing sensitive data, having Non-IID (Independent and Identically Distributed) data, and with non trusted stakeholders or clients.}
    
    \item A \addtxtRev{security} framework that uses FL to detect, in a privacy preserving fashion, cyberattacks affecting IoT devices. The proposed framework covers both anomaly detection and classification approaches using multi-layer perceptron and autoencoder \addtxtRev{neural network} architectures.
    
    \item A pool of experiments measuring the performance of the proposed framework when detecting malware in IoT devices. \changeRev{After that, the following approaches have been compared}{To that end, the next scenarios have been compared}: i) a centralized approach where all the data is shared, ii) a distributed approach where each entity trains an independent model with its local data, and iii) a federated approach where a joint model is generated sharing the local model updates. Two different federated learning algorithms that differ in the number of communications \addtxtRev{(model sharing updates)} with the server have been considered in the previous comparison.
    
    \item The evaluation of the impact of several adversarial attacks affecting our FL solution. The objective is to measure how the federated models performance degrades when some clients are malicious \addtxtRev{and send tampered model updates}. Besides, it has been evaluated how different aggregation functions acting as countermeasure mechanisms improve the model resilience against adversarial attacks.
    
    \item \addtxt{The discussion on the adversarial results, the communication and computation costs and the design of the framework, describing possible issues and drawbacks \addtxtRev{in B5G scenarios}, together with their possible solution.}
\end{itemize}


The remainder of this paper is organized as follows. Section \ref{sec:related} describes related work on AI for IoT cybersecurity, FL algorithms, vulnerabilities and countermeasures, and datasets containing IoT cyberattack data. Section \ref{sec:use_case} depicts an IoT scenario with privacy requirements that are accomplished by the N-BaIoT dataset, which serves as use case for this work. Section \ref{sec:architecture} details the design and implementation of the proposed framework, which uses FL to detect malware affecting IoT devices. Section \ref{sec:attacks} defines the adversarial attacks and countermeasures tested against the proposed framework. Section \ref{sec:experiments} shows the results of the experiments done in this work, comparing the federated approaches against traditional ones and detailing the results of the adversarial settings. \addtxt{Section \ref{sec:discussion} analyzes the lessons learned as well as the possible drawbacks of the architecture.} Finally, Section~\ref{sec:conclusion} shows the conclusions of this research and future directions.

\section{Related work}
\label{sec:related}

This section details the current state-of-the-art in different topics covered in the present work. First, it reviews the usage of AI for IoT cybersecurity, with special consideration of FL. Then, it describes the main literature on adversarial attacks against the FL process and their possible mitigations. Finally, it reviews the available datasets modeling cyberattacks on IoT devices.

\subsection{Artificial Intelligence for IoT Cybersecurity}

Traditional AI techniques have been widely applied in the literature to detect cybersecurity issues in IoT scenarios. In \cite{device_behavior}, existing works on device behavior fingerprinting were surveyed, including those targeting IoT security. This work shows how IoT security solutions are turning nowadays towards the application of ML and DL techniques. In this direction, the authors of \cite{edima} used ML techniques for early detection of heterogeneous malware affecting IoT devices. Another work was presented in \cite{iot_intrusion_survey}, where many intrusion detection systems for IoT devices were reviewed, providing recommendations for designing robust and lightweight intrusion detection solutions for IoT.

In the last years, FL is gaining importance in the field of cybersecurity, with several works already using this paradigm for IoT security. In this context, the research proposed in \cite{diot} clearly stated the data privacy problem of traditional AI-based solutions, but the evaluation took place on a private dataset. Also, the data was randomly split among clients, which can be improbable in realistic scenarios, as the one considered in this work, in which each client data comes from a different distribution in general.
The works presented in \cite{fed_iiot,fl_iiot} also have very similar objectives, but these researches were conducted specifically for industrial IoT devices, and they analyzed respectively application samples and sensor readings rather than network data, as we do in this work. In \cite{blockchain_fl_id}, FL was studied through the use case of intrusion detection systems. This work also includes blockchain technology to mitigate the problems faced in adversarial FL. However, it concentrates on the early steps of intrusion detection rather than detecting already running malware, and it does not focus specifically on IoT devices.

In summary, this section has shown the lack of solutions dealing with FL approaches considering data generated by decentralized sources to detect malware affecting IoT devices and scenarios.

\subsection{Federated Learning Algorithms, Vulnerabilities and Countermeasures}

Focusing on FL algorithms and their particularities, the work of \cite{fedavg} defines the term \textit{federated learning} by characterizing the decentralized non-IID \rmvtxtRev{(Independent and Identically Distributed) }optimization problem. They propose the Federated Averaging (FedAVG) algorithm that now serves as a powerful baseline for many researches using FL. In this algorithm, several clients use their individual datasets to collaboratively train a global model, thanks to the coordination provided by a central server. The role of the server is to average the parameters of the models sent by the clients and return the resulting global model to them. This process is iterated until a terminating condition is met. FL has matured a lot since then and several surveys (\cite{fl_advances, fl_concepts}) review the latest advances in that domain.

Because of its decentralized nature, FL shares the threat among multiple entities, namely the clients and the server. The work of \cite{adversarial_fl_survey} reviews many of the problems that can arise when considering an adversarial setup, as well as most of the well-known defenses to protect the system against that. In \cite{data_poisoning}, several data poisoning attacks against Support Vector Machines were defined. Their baseline experiment used the idea of label flipping, in which the binary label of some datapoints in the training set is inverted to hinder the training of the model. In \cite{krum}, the authors studied the resilience of a distributed implementation of Stochastic Gradient Descent against arbitrarily behaving (Byzantine) adversaries. To that end, a model poisoning attack from the standpoint of a malicious client, that is capable of estimating the gradient, was experimented. First, it demonstrated that the usual model averaging step executed by the server in most FL algorithms does not handle even a single malicious client in the federation. More generally, they proved that no model aggregation function, linear in the models sent by the clients, is robust against Byzantine adversaries. In \cite{cm_tm}, two additional robust model aggregation functions were proposed. In particular, they are based on the coordinate-wise median and the coordinate-wise trimmed mean of the models sent by the clients to the server. The authors of \cite{s-resampling} proposed resampling to reduce heterogeneity in the distribution of the models sent by the clients. It is meant to be applied before using a robust aggregation function, and it aims at reducing the side-effects that such a function has when applied to models trained with non-IID datasets.

\addtxt{Finally, in the literature there are also decentralized algorithms for securing distributed computing tasks which use reinforcement learning approaches in scenarios where there is no direct trust in the clients but a correct result is desired even if there are some clients with malicious behavior \cite{christoforou2013applying}.}

\subsection{Datasets Modeling Cyberattacks Affecting IoT Devices}

Datasets are key for AI in general and FL in particular. In this sense, several public network datasets about IoT security can be found in the literature. \tablename~\ref{tab:datasets} reviews some of the most interesting ones. All of those datasets are generated at a central location, but for some of them, a realistic splitting strategy is doable to let them be used in FL approaches. In this context, the \textit{Splitting} column presents our proposal in terms of possible strategies to split the dataset among different entities. In the \textit{Device} splitting strategy, the dataset already has the traffic from each device placed into a different file. The \textit{IP} strategy would consist of grouping the dataset samples by IP address to manually isolate the traffic of each device. The \textit{Scenario} splitting strategy would take advantage of the fact that the dataset was generated in several different scenarios, and it might be possible to consider each scenario as coming from a different client. Finally, the \textit{Unrealistic} label means that no realistic (non-IID) strategy was found to make the dataset appear to come from several sources.

In \cite{n-baiot}, a dataset called N-BaIoT was produced by preprocessing the traffic generated by 9 commercial IoT devices of various types, either infected by Mirai or BASHLITE (two botnet malware attacks), or uncorrupted. In \cite{medbiot}, a medium-sized network of 83 real or emulated IoT devices is considered to produce the MedBIoT dataset. It uses the same packet preprocessing as in N-BaIoT, but here other stages of malware traffic are considered (infection, propagation and communication with the command and control server). In \cite{kitsune}, the evaluation dataset consists of a network made of 8 security cameras suffering from several attacks. Additionally, they included another network consisting of 9 commercial IoT devices, among which one was infected by Mirai. \cite{bot_iot} proposes a dataset called Bot\_IoT, that contains legitimate and simulated IoT network traffic, including different attacks. The dataset TON\_IoT \cite{ton_iot} consists of heterogeneous data sources (network data but also sensor readings, operating system logs and telemetry data) about a network containing several IoT/IIoT devices. In \cite{iot_ba_traces}, the authors propose a dataset collecting benign and volumetric attacks traffic traces for 27 IoT devices. The main purpose of this dataset is to evaluate volumetric attacks perpetrated against a network containing real commercial IoT devices. The dataset proposed in \cite{iot_network_intrusion} was generated with the traffic of 2 home IoT devices under multiple attack scenarios. It also includes simulated Mirai traffic appearing to come from the IoT devices. Finally, IoT-23 \cite{iot_23} is a dataset consisting of 20 captures that include malware activity as well as 3 captures of benign IoT traffic. 

To conclude, it is worthy to mention that there is a lack of dataset suitable for FL approaches detecting malware in IoT devices. Existing FL-based solutions must consider split centralized datasets in order to apply federated techniques.

\begin{table}[ht]
\centering
\begin{center}
\setcellgapes{\cellgps}
\makegapedcells
\begin{tabular}{clcl}
\thline
\textbf{Ref.} & \textbf{Name} & \textbf{Year} & \textbf{Splitting} \\ \mhline
\cite{n-baiot} & N-BaIoT & 2018 & Device\\ \hline
\cite{medbiot} & MedBIoT & 2020 & IP \\ \hline
\cite{kitsune} & Kitsune & 2019 & Unrealistic \\ \hline
\cite{bot_iot} & Bot\_IoT & 2018 & IP, scenario \\ \hline
\cite{ton_iot} & TON\_IoT & 2019 & IP, scenario \\ \hline
\cite{iot_ba_traces} & IoT benign \& attack traces & 2019 & IP, scenario \\ \hline
\cite{iot_network_intrusion} & IoT network intrusion & 2019 & Unrealistic \\ \hline
\cite{iot_23} & IoT-23 & 2020 & Unrealistic \\ \hline
\hline
\end{tabular}
\caption{Public IoT network datasets.}
\label{tab:datasets}
\end{center}
\end{table}

\section{Use Case: IoT Scenario Affected by Malware} 
\label{sec:use_case}

This section presents the characteristics of the scenario defined in this work and explains the details of the dataset used to evaluate the performance of the proposed framework. 

\change{Our cities have millions of IoT devices sensing heterogeneous pieces of data which are connected to the internet}{Our cities have millions of IoT devices connected to the Internet and sensing heterogeneous pieces of data. The number and heterogeneity of devices will increase exponentially with the advent of B5G networks,} \addtxtRev{as they enable new verticals and scenarios based on the enhanced network performance in terms of latency and throughput \cite{samdanis2020road}. Some examples are Unmanned Aerial Vehicle Services, Holographic Teleportation, or Extended Reality.} \change{Some of these pieces}{In such a context, privacy issues frequently appear when pieces of sensed data} belong to sensitive aspects of our daily lives or organizations \addtxtRev{\cite{fadlullah2020hcp}}. As demonstrated, IoT devices are constrained in terms of resources and have not been designed with security in mind, making them vulnerable to a wide variety of malware. In these scenarios, traditional AI\addtxt{-based detection} approaches are not suitable due to the impossibility of training centralized models with sensitive data belonging to different organizations or subjects. Because of that, FL is raising as a key mechanism to detect anomalous behaviors and trigger mitigation mechanisms \addtxt{in privacy-sensitive scenarios enabled by 5G and B5G networks}. However, FL also suffers from inherent problems of dealing with unknown and, therefore, untrusted parties. Malicious clients executing poisoning attacks over data and models is one of the best examples in this direction. Following the previous characteristics, this work considers the following key aspects for the defined scenario: i) data is non identically distributed across the IoT devices (owned by the clients), ii) it is needed to detect anomalies provoked by unseen or zero-day malware affecting IoT devices, iii) it is required to classify well-known malware affecting different IoT devices, iv) adversaries can be present among the federated clients, so some countermeasures should be applied. 

\change{Taking into account the previous requirements, we analyzed the existing datasets and selected the most suitable one. At this point, it is also worthy of mentioning that even though some public IoT malware datasets exist, one important requirement that we considered was the fact that the dataset should be split into several parts to perfectly fit into our scenario. Among the reviewed datasets, N-BaIoT is the most suitable for that kind of splitting. Specifically, the dataset already separates the devices into different files. Since malware detection is also the focus of this dataset, it constitutes the ideal candidate, and it was therefore selected to evaluate our approach.}
{Several public datasets aligned with B5G application scenarios and IoT malware exist in the literature. Among them, N-BaIoT is the most suitable to evaluate privacy-preserving collaborative training. Specifically, this dataset already separates the IoT devices' traffic data into different files, making it easy to split it into several non identically distributed parts for a realistic federated setting. For that reason, we selected N-BaIoT to evaluate our approach. Note that a drawback of this dataset is that it only contains the data from 9 IoT devices, which is a limitation for the experiments as it limits the maximum number of clients that can be considered.}

N-BaIoT contains the preprocessed packets from the traffic of \addtxt{the} 9 IoT devices. All devices have generated some traffic while non-corrupted (benign samples) and while being infected by Mirai and BASHLITE. All devices, except the Ennio doorbell and the webcam, also have generated some traffic while infected by Mirai. \tablename~\ref{tab:devices} shows the number of benign and attack samples for each device, as well as the total.

\newcommand{\samples}[2]{\makecell[l]{\numprint{#1}}}
\newcommand{\totalsamples}[2]{\textbf{\makecell[l]{\numprint{#1}\\(#2\%)}}}

\begin{table}[ht!]
\begin{center}
\setcellgapes{\cellgps}
\makegapedcells
\begin{tabular}{lll}
\thline
\textbf{Device} & \textbf{\makecell[l]{Benign\\ samples}} & \textbf{\makecell[l]{Attack\\samples}} \\ \mhline
Danmini Doorbell & \samples{49548}{4.87} & \samples{968750}{95.13} \\ \hline
Ecobee Thermostat & \samples{13113}{1.57} & \samples{822763}{98.43} \\ \hline
Ennio Doorbell & \samples{39100}{11.00} & \samples{316400}{89.00} \\ \hline
\makecell[l]{Philips B120N10\\ Baby Monitor} & \samples{175240}{15.95} & \samples{923437}{84.05} \\ \hline
\makecell[l]{Provision PT-737E\\ Security Camera} & \samples{62154}{7.50} & \samples{766106}{92.50} \\ \hline
\makecell[l]{Provision PT-838\\ Security Camera} & \samples{98514}{11.77} & \samples{738377}{88.23} \\ \hline
\makecell[l]{Samsung SNH-1011-N\\ Webcam} & \samples{52150}{13.90} & \samples{323072}{86.10} \\ \hline
\makecell[l]{SimpleHome XCS7-1002\\-WHT Security Camera} & \samples{46585}{5.40} & \samples{816471}{94.60} \\ \hline
\makecell[l]{SimpleHome XCS7-1003\\-WHT Security Camera} & \samples{19528}{2.30} & \samples{831298}{97.70} \\ \mhline
\textbf{Total} & \totalsamples{555932}{7.87} & \totalsamples{6506674}{92.13} \\ 
\hline
\hline
\end{tabular}
\caption{Number of benign and attack samples for each device.}
\label{tab:devices}
\end{center}
\end{table}

Each sample in the dataset corresponds to a network packet sniffed by Wireshark. For each, 115 numerical features characterizing the context of the packet were extracted. \change{For example, one feature is the mean packet size over the last 10 seconds in the traffic between the current packet source IP and destination IP.}{The available features are statistics about the size, count and jitter of aggregated network packets, in the last 100 ms, 500 ms, 1.5 sec, 10 sec and 1 min. For example, one feature is the mean packet size over the last 10 seconds in the traffic between the current packet source IP and destination IP.} Noticeably, the features of packets captured in a very short time interval are highly correlated. This means that this dataset needs to be handled with care in order to reduce as much as possible the data leak between the train and the test sets when separating a given file into those two parts. To that end, we always used chronological splitting to make the train and test parts, and we left a small set of samples unused between the train part and the test part for each file in the dataset.


After analyzing the most relevant characteristics of the dataset, we also reviewed some of the most notable existing works on N-BaIoT \cite{n-baiot, n-baiot_sparse_repr, n-baiot_unsupervised_feature_reduction, n-baiot_supervised_feature_reduction, n-baiot_collaborative, n-baiot_unsupervised_grey_wolf}. Most of those focus on unsupervised anomaly-detection solutions, using only the benign part of the dataset to train. Still, in \cite{n-baiot_unsupervised_grey_wolf}, the hyper-parameters are tuned using also some attack data, and in \cite{n-baiot_supervised_feature_reduction}, a supervised classification is considered instead. Some works use multiple samples in order to detect potential malware, and others focus on the more granular task of single-sample classification. In our methodology, both the supervised and the unsupervised situations are considered, and the attention is placed on single-sample analysis. \addtxt{In the supervised situation, as N-BaIoT contains 10 different attacks performed using Mirai and BASHLITE, we use all the available attacks labeled using the same class (attack) in order to detect as many attacks as possible.} Note that in \cite{n-baiot_collaborative}, a collaborative learning approach is proposed. However, the assumed scenario and the goal are different from ours, as they focus on building one model per device with the assumption that the data of a single device comes from 2 or 3 different sources.

\section{Federated Learning-based Framework and Deployment} %
\label{sec:architecture}

This section details the architectural design of the proposed FL-based framework, describing its components and how they interact with each other during the model training and evaluation processes. Besides, it also depicts how the framework is deployed for our validation use case, which leverages the N-BaIoT dataset.

\newcommand{\alldevices}{\mathcal{S}}
\newcommand{\allbut}[1]{$\alldevices - \{$#1$\}$}
\newcommand{\singleton}[1]{$\{$#1$\}$}

\newcommand{\nfeatures}{115}
\newcommand{\nepochs}{E}
\newcommand{\bs}{B}
\newcommand{\stepsize}{S}
\newcommand{\dktrain}{\mathcal{D}_{k}^{\textit{Train}}}

\newcommand{\dkthr}{\mathcal{D}_{k}^{\textit{Thr}}}
\newcommand{\allk}{\hspace{0.1cm}\forall k \in [K]}
\newcommand{\thrkmeanstd}{thr_{k} = mean(\msekthr) + std(\msekthr)}

\newcommand{\pk}{\mathcal{P}_{k}}
\newcommand{\allkk}{{\hspace{0.2cm}\forall k \in [K]}}

\newcommand{\nrounds}{T}
\newcommand{\nbatchesk}{|\mathcal{B}|}
\newcommand{\minibavg}{\textsc{Mini-batch avg}}
\newcommand{\minibagg}{\textsc{Mini-batch aggregation}}
\newcommand{\multieavg}{\textsc{Multi-epoch avg}}
\newcommand{\multieagg}{\textsc{Multi-epoch aggregation}}

\newcommand{\alllf}{\textit{all labels flipping}}
\newcommand{\modelc}{\textit{model cancelling}}
\newcommand{\gradientf}{\textit{gradient factor}}

\begin{figure}[ht!]
\centering
\includegraphics[width=.95\linewidth]{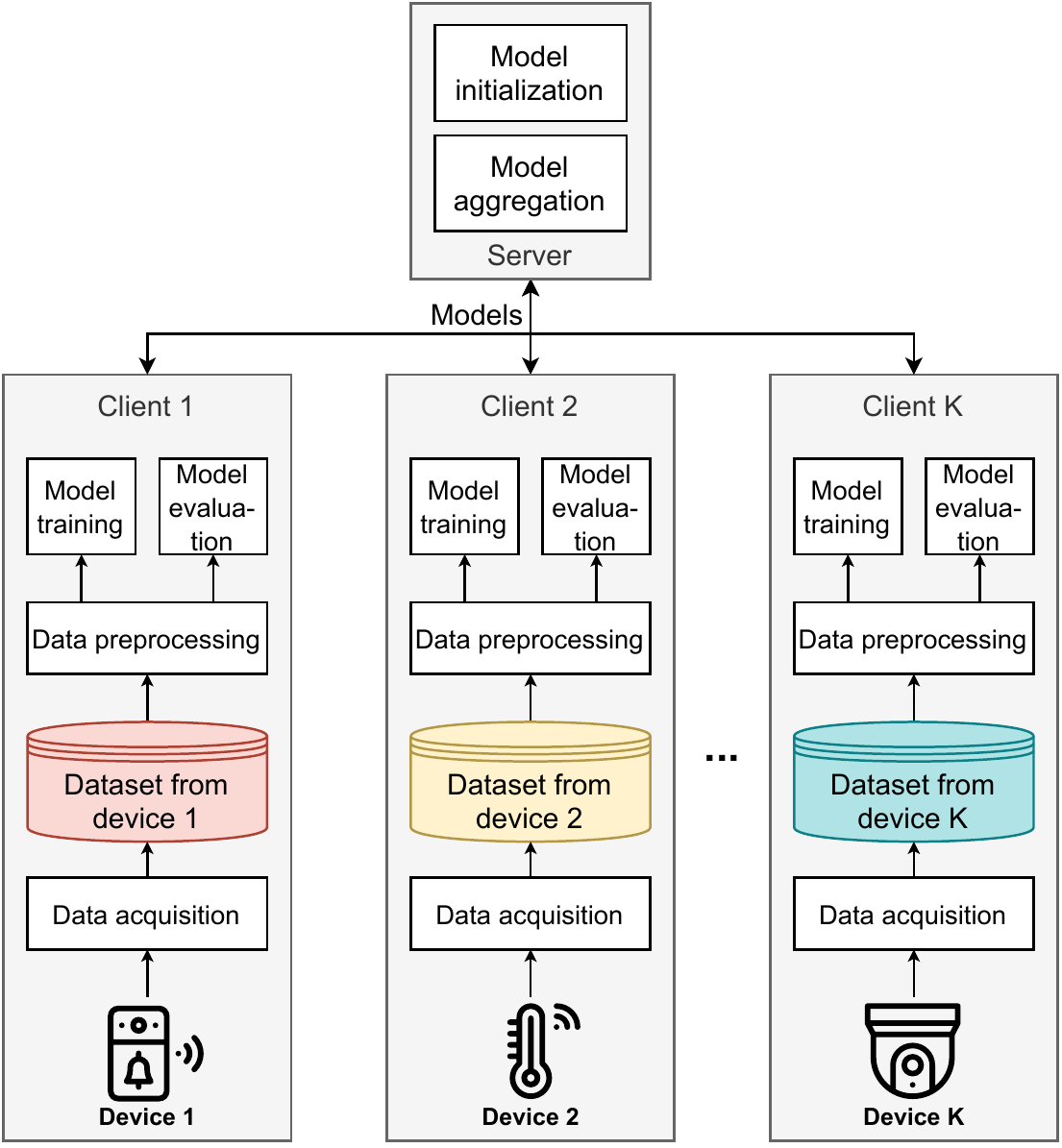}
\caption{Framework architecture and its components. The sharing of normalization values and the collaborative hyper-parameter selection are omitted for simplicity.}
\label{fig:system_architecture}
\end{figure}

The framework architecture, depicted in \figurename~\ref{fig:system_architecture}, consists of $K$ clients that own the data from a single device each and a server that coordinates the FL process. The following sections provide the design details about each component making up the proposed architecture. The code used to implement the whole pipeline is available at \cite{valerian_code}.

\subsection{Client}
\begin{figure}[ht!]
\centering
\includegraphics[width=.95\linewidth]{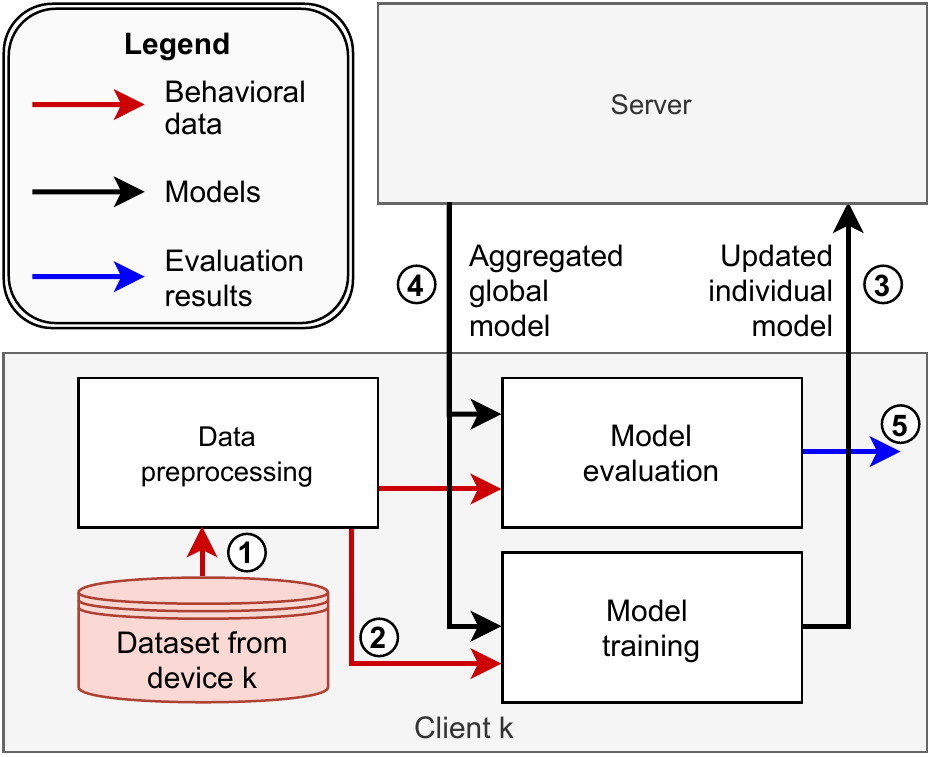}
\caption{Detailed view of the client architecture during training and evaluation. The initial model sharing (by the server) is omitted for simplicity. Steps 1, 2, 3 and 4 are meant to be repeated several times before the model is evaluated (step 5).}
\label{fig:client_architecture}
\end{figure}

Considering that IoT devices generally have limited resources and modest reliability, the clients in charge of training the models are not the devices to be protected, but other entities capable of collecting the traffic of the IoT devices present in the same network\addtxt{, such as B5G base stations or other access points}. \addtxtRev{In this sense, in the B5G architecture \cite{dogra2020survey}, the present system would be incorporated in the RAN SLICING Edge Nodes or in the CLOUD SLICING Fog Nodes.} This system falls into the category of cross-silo FL, as defined in \cite{fl_advances}, where the federated clients are few but powerful and reliable. Note that each client can own several IoT devices, but for the sake of simplicity, the architecture and the experiments are described with a single one per client. \figurename~\ref{fig:client_architecture} details the architecture of a client after data acquisition, as well as its interactions with the server. The dataset and the components depicted in the figure mentioned above are explained in detail in the remainder of this section.

\subsubsection{Data Acquisition}
The client is in charge of gathering the traffic data from the device under observation. This can be done, for example, by using port mirroring on the switch that connects the IoT device (as described in \cite{n-baiot}). In our solution, since we use an existing dataset, this component was not developed.

\subsubsection{Dataset}
\label{sec:dataset_splitting}
Two situations are considered here. The supervised situation assumes a setup in which each client has access to labeled data from its own device. In the second situation, we assume that each client only has access to the benign traffic of its device. Since getting a large quantity of benign traffic data is generally easy and does not need manual labelling, this situation is often termed as unsupervised in the literature \cite{n-baiot, n-baiot_unsupervised_feature_reduction, n-baiot_unsupervised_grey_wolf}. However, in the strict sense of the term, it refers to single-class supervised learning \cite{n-baiot_unsupervised_feature_reduction}.

Note that in reality, a supervised situation with several clients able to generate a decent amount of labeled data is plausible but uncommon. Therefore, the supervised solution has two main motivations. The first one is to have a comparison point for the unsupervised solution. As the supervised situation is easier to tackle and is more controllable than the unsupervised one, the second motivation is to be able to go as in-depth as possible in our experiments, to potentially reveal vulnerabilities or other concerns about using FL for malware detection.

\begin{figure}
\centering
\begin{subfigure}{.2\textwidth}
  \raggedright
  \includegraphics[width=.90\linewidth]{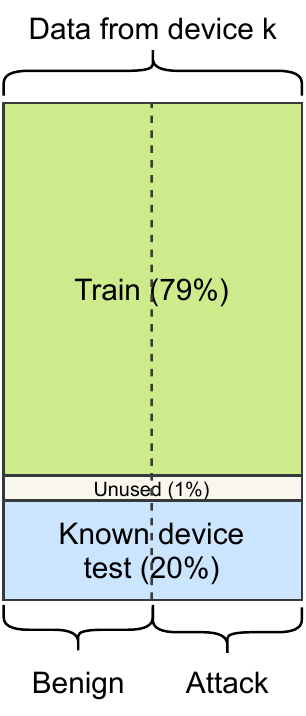}
  \caption{Supervised approach.}
\end{subfigure}%
\begin{subfigure}{.2\textwidth}
  \raggedleft
  \includegraphics[width=.90\linewidth]{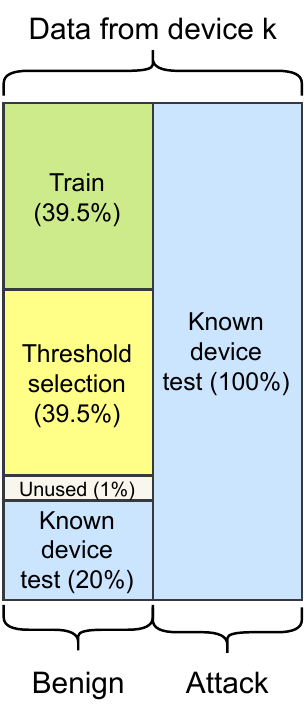}
  \caption{Unsupervised approach.}
\end{subfigure}
\caption{Initial splitting of the dataset owned by client k for the supervised and the unsupervised situations. The relative size of the benign part with respect to the attack part is not respected for readability.}
\label{fig:dataset_splitting}
\end{figure}

\figurename~\ref{fig:dataset_splitting} shows for both situations how we have split the dataset of a single device for training and testing purposes. In the supervised situation, each dataset is split chronologically between 3 parts: the train set (79\%), the aforementioned unused set (1\%) and the test set (20\%). In the unsupervised situation, only the benign data is available for training, so this part is split between 4 different sets: the train set (39.5\%), a so-called threshold-selection set (39.5\%), the unused set (1\%) and the benign part of the test set (20\%). All attack data is available for the final testing of our experiments.

\change{We also decided to re-balance the dataset in three different ways to reach the following class proportions for every device:}{We also decided to re-balance the dataset in three different ways in order to cover several possible data scenarios. We selected the following class proportions for every device:}

\begin{itemize}

    \item 7.87\% benign traffic and 92.13\% attack traffic. This is the original dataset balance, with the difference that the proportion of each class now does not vary across the devices.

    \item 50\% benign traffic and 50\% attack traffic, making the dataset perfectly balanced for binary classification.
    
    \item 95\% benign traffic and 5\% attack traffic. This is much more representative and aligned with the reality\addtxt{, where usually much more benign data is available}.
    
\end{itemize}

It is important to note that these three re-balancings lead to three different problems. Both train and test sets are indeed affected by each change, and the goal is not to compare the impact on the model performance when re-balancing the classes. This is an operation to make the results as broad as possible rather than a way to handle the dataset imbalance.

The number of samples per device is also fixed to a constant in order to make the results less dependent on the number of training instances and to keep the dataset size fixed no matter what the proportions of classes are. \numprint{100000} samples per device are used for the supervised solution and only \numprint{10000} for the unsupervised one. This is because the unsupervised training takes more time to converge, and it only needs benign data (which rarely reaches \numprint{100000} samples anyway) in its train set. 
\change{If more samples than available are needed to reach the desired number of samples of a specific class, upsampling is used (repeating some samples multiple times). Otherwise, downsampling is performed, keeping only a subset of the available samples.}{The procedure followed to reach at the same time these numbers of samples and the desired class proportions is to use upsampling (duplicating the original samples) when more samples than available are needed, and downsampling (keeping only a subset of the original samples) otherwise.}
Either way, it takes place after splitting the data between the train and test sets, so no data leak is created. After setting the proportions of each class and the desired number of samples per device, we obtain a dataset where the number of samples and the proportions of classes are the same for all devices.

After this balancing process, the train set of client $k$ is defined as $\dktrain$ and its threshold selection set (in the unsupervised solution) is defined as $\dkthr$. The number of training samples of client $k$ is $n_{k} \coloneqq |\dktrain|$.

\newcommand{\xmax}{x^{max}}
\newcommand{\xmin}{x^{min}}
\newcommand{\xkmax}{x_{k}^{max}}
\newcommand{\xkmin}{x_{k}^{min}}

\subsubsection{Data Preprocessing}
This component is in charge of normalizing the samples. Min-max feature scaling is used, \ie\ $x' = \frac{x - \xmin}{\xmax - \xmin} \in \mathbb{R}^{115}$, where operations are applied element-wise. The normalization values are computed only with the train set that the client owns. Note that each client $k$ originally has its own normalization values $\xkmin$ and $\xkmax$. As we will see in Section \ref{subsubsection_collaborative_normalization}, clients can collaborate in order to know the global values for $\xmin$ and $\xmax$, over the train sets of all devices.

\newcommand{\wk}{w_{k}}
\newcommand{\wi}{w^{(i)}}
\newcommand{\wki}{w_{k}^{(i)}}
\newcommand{\msekthr}{\textit{MSE}(\dkthr ; \wk)}

\subsubsection{Model training}
\label{subsubsection_model_training}
The purpose of the component is to train the federated ML model that will be used for malware detection. To that end, we first present the architectures used for classification and anomaly-detection. Later, for two different FL algorithms, we explain how this component interacts with the server. Throughout the rest of this work, the model parameters of client $k$ are referred as $\wk$ and the global model parameters as $w$. Further, with $d$ the number of dimensions of $w$ and with $i \in [d]$, $\wi$ specifies the $i$\textsuperscript{th} dimension of $w$. \addtxt{Note that most well-known ML models are compatible with our framework, as long as the trained models do not vary in structure among clients and do not store training data explicitly (otherwise sharing them would compromise privacy).}

\paragraph{Supervised situation.}
In this setup, a binary classification task is considered with four different architectures of multi-layer perceptrons (MLP) with 1 output neuron:

\begin{itemize}
    \item Classifier A: No hidden layer (linear model).
    \item Classifier B: 1 hidden layer with 115 hidden neurons.
    \item Classifier C: 2 hidden layers with 115 and 58 hidden neurons, respectively.
    \item Classifier D: 3 hidden layers with 115, 58 and 29 hidden neurons, respectively.
\end{itemize}
After each hidden layer, the exponential linear unit (ELU) \cite{elu} activation function is used.
Note that the numbers of hidden neurons that are tried, 115, 58 and 29, correspond respectively to 100\%, 50\%, and 25\% of the input dimension.

\paragraph{Unsupervised situation.}
In this setup, autoencoders are used for anomaly detection, following a similar methodology as the authors of N-BaIoT \cite{n-baiot}. An autoencoder is a special form of feed-forward neural network made of two parts, the encoder and the decoder. The encoder transforms the input by reducing its number of dimensions to a value defined as the coding dimension, and the decoder tries to map the encoded input back to the original input. It is trained by minimizing the Mean Squared Error (MSE) between the reconstructed features and the input. In order to use this principle for anomaly detection, an autoencoder is trained with benign data, learning how to reconstruct it, such that it has a low reconstruction error on future benign data and a high reconstruction error on anything that deviates from benign data. Once the training process is completed, a threshold is set based on statistics about the reconstruction error of benign data. During testing, if a sample has a mean squared reconstruction error higher than the specified threshold, it is considered as anomalous (positive), otherwise it is considered as benign (negative). The threshold formula used in this work comes from \cite{n-baiot}, and selects for client $k \in [K]$ the threshold 
\begin{equation}
    \thrkmeanstd
\label{eq_threshold_mean_std}
\end{equation}
where $\textit{MSE}(\,\cdot\,; \wk)$ is the mean squared reconstruction error computed with model parameters $\wk$.
Multiple autoencoder architectures are investigated during the grid searches.
\begin{itemize}
    \item Autoencoder A: 1 hidden layer of 29 neurons (shallow autoencoder).
    \item Autoencoder B: 3 hidden layers of 58, 29 and 58 neurons.
    \item Autoencoder C: 7 hidden layers of 86, 58, 38, 29, 38, 58 and 86 neurons. These numbers of neurons and layers correspond roughly to those used in the solution of the creators of N-BaIoT \cite{n-baiot}.
\end{itemize}

Once again, ELU is used after each hidden layer. All considered architectures have 29 coding dimensions. A low number of dimensions is a way to constrain the autoencoder to learn a representation that is more specific to benign data. Indeed, using 115 coding dimensions would make the autoencoder able to learn the identity function for any input vector in $\mathbb{R}^{115}$, making it produce low reconstruction errors for very unusual data, even if it is trained only with benign data. The choice of 29 coding dimensions corresponds to what is used in \cite{n-baiot}. Without looking at labeled data, it is hard to make a better selection of this hyper-parameter.
The numbers 86, 58, 38 and 29, correspond respectively to 75\%, 50\%, 33\% and 25\% of the input dimension.

\paragraph{Interactions with the server.}
Two FL algorithms, \minibagg\ and \multieagg, deriving from the popular FedAVG \cite{fedavg} are considered. For both algorithms, the main difference with FedAVG is that we consider the aggregation function as a parameter of the algorithm. Therefore, the server can try other aggregation functions than averaging. Also, it provides a more practical implementation, where the learning rate varies over the training process.

In \minibagg, the Model Training component trains the model with a single mini-batch of data before sending it to the server for aggregation. The Model Training component then receives the new aggregated global model, with which the training can continue. This process is repeated until a number $E$ of epochs over the full train set are completed.
In \multieagg, the model is trained for all $E$ epochs at once before being sent to the server for aggregation.
A potential drawback is that, as explained in \cite{fedavg}, averaging models could have arbitrarily bad results because of the non-convexity of the objective. This problem is much more likely for \multieagg\ as the models are trained separately for much longer before being aggregated. In order to try to mitigate that, the training of \multieagg\ is repeated for $T = 30$ rounds, with a learning rate decreasing over the rounds.

\rmvtxt{Besides, it is worthy of mentioning that MINI-BATCH AGGREGATION has much higher communication costs than MULTI-EPOCH AGGREGATION at i.e. $E \cdot \frac{n_{k}}{B}$ model transmissions per client for the full training, where $B$ is the batch size. Note that in terms of computation cost, it also indicates the number of local model updates performed by each client. On the other hand, MULTI-EPOCH AGGREGATION just requires the clients to transmit the model to the server once per round, for a total of $T$ transmissions per client. However, the number of local model updates is also $T$ times larger, i.e. $T \cdot E \cdot \frac{n_{k}}{B}$ for client $k$.}

\subsubsection{Model Evaluation}
After the model has been trained for a satisfying number of iterations through the FL process, it is ready to be evaluated. In order to assess the robustness of the trained models, we evaluate them on different test sets. The \textit{known devices} performance is given by the evaluation of the model on the test part of the data from the devices owned by the clients. The \textit{new device} performance is computed on the data from a device that is totally new to all clients in the federation (it has not been seen during the selection of the normalization values, the hyper-parameter selection nor the training). Note that the new device's test set thus has a different distribution than the training sets in general.

\subsection{Server}
In the proposed framework architecture, the server is in charge of coordinating the training efforts of the federated clients. Specifically, it initializes the model at the very beginning, and it aggregates the models sent by the clients into a so-called global model. It also has to coordinate the additional steps described in Section \ref{subsection_additional_concerns}, \ie\ the collaborative normalization, the collaborative grid searches, and the collaborative threshold selection (for the anomaly-detection approach). \addtxtRev{In the B5G architecture context \cite{dogra2020survey}, the server component would be placed in the CLOUD SLICING layer, either on the Fog Nodes or in the Cloud Data Centres, depending on the scope of the clients covered.}

\subsubsection{Model Initialization}
The server is in charge of initializing the weights of the initial model. Once it is done, the initial model is shared with all clients, and the training process can start. It is worth noting that each client starts with the same model.

\subsubsection{Model Aggregation}

After receiving the updated model parameters of each client \ie\ $\{\wk:\allk\}$, the server has to aggregate them to form the new global model parameters $w$. With the baseline averaging approach, the formula for this is given by $w \coloneqq \sum_{k=1}^{K}\frac{1}{K}\wk$. When this aggregation function is used, we refer to the algorithms \minibagg\ and \multieagg\ as \minibavg\ and \multieavg, respectively. Note that a weighted averaging could be used if the number of samples varied among clients. Other aggregation functions can also be used to provide additional security, as indicated in Section  \ref{subsection_robust_aggregation}.

\addtxt{Although there is a server in charge of model aggregation in the current design of the framework due to the advantages of having an entity coordinating the process, it would be possible to move the \textit{Model Initialization} and \textit{Model Aggregation} steps into the clients themselves, or decentralizing the server into several entities. For this, Blockchain technologies would be used as a decentralized database where each client would share its local model and retrieve the models of other clients when performing the aggregation. Thus, the framework would be totally decentralized without an entity coordinating the generated models. }

\subsection{Additional Concerns of the Proposed Framework}
\label{subsection_additional_concerns}

This section summarizes some additional concerns that arise when performing the usual full pipeline of ML in a federated way. Specifically, the steps of normalization and hyper-parameter selection must be given attention. Besides, for the unsupervised solution, the step of threshold selection requires special considerations as well.

\subsubsection{Collaborative Normalization}
\label{subsubsection_collaborative_normalization}

Since min-max feature scaling is used, each client $k$ can compute $\xkmin  \in \mathbb{R}^{\nfeatures}$ and $\xkmax \in \mathbb{R}^{\nfeatures}$ locally and the server can compute the global minimum and maximum $\xmin \in \mathbb{R}^{\nfeatures}$ and $\xmax \in \mathbb{R}^{\nfeatures}$ as the element-wise minimum and maximum, respectively, of those values. This procedure is detailed in algorithm \ref{algorithm_collaborative_normalization}. Note that it gives the global minimum and maximum as if they were computed directly on the combination of the train sets of all clients. This has the drawback of requiring each client to leak its exact values of minimum and maximum for each of the 115 features.

\begin{algorithm}[ht!]
\SetAlgoLined
\SetAlgoNoEnd
\SetKwInOut{Input}{input}
\SetKwBlock{Begin}{}{}
\SetKw{Parallel}{in parallel} 
\textbf{Server executes:}
\Begin{
    \For{each client $k \in [K]$ \Parallel}{
        $\xkmin, \xkmax \leftarrow$ ClientMinMax$(k)$
    }
    $\xmin \leftarrow \min_{k \in [K]}\{\xkmin\}$\\
    $\xmax \leftarrow \max_{k \in [K]}\{\xkmax\}$\\
    \For{each client $k \in [K]$ \Parallel}{
        ClientStoreMinMax$(k, \xmin, \xmax)$
    }
}

\BlankLine
\BlankLine
\BlankLine

\textbf{ClientMinMax}$(k)$: \tcp*[h]{Run on client $k$}
\Begin{
    $\xkmin$ = $\min\{\dktrain\}$\\
    $\xkmax$ = $\max\{\dktrain\}$\\
    return $\xkmin, \xkmax$ to server
}

\BlankLine
\BlankLine
\BlankLine

\textbf{ClientStoreMinMax}$(k, \xmin, \xmax)$ \tcp*[h]{Run on client $k$}
\Begin{
    Store $\xmin$ \tcp*[h]{Client $k$ now has access to $\xmin$}\\
    Store $\xmax$ \tcp*[h]{Client $k$ now has access to $\xmax$}
}
\caption{\textsc{Collaborative normalization}. $[K]$ is the set of clients and $\dktrain$ is the set of datapoints used by client $k$ for training; min and max are the element-wise minimum and maximum. Since they are always applied with vectors in $\mathbb{R}^{115}$, they also output a value in $\mathbb{R}^{115}$.}
\label{algorithm_collaborative_normalization}
\end{algorithm}

\subsubsection{Collaborative Grid Search}
\label{subsubsection_collaborative_gs}
Two types of hyper-parameters should be distinguished: the ones that need to be common to every client, mainly about the architecture of the model (number of layers, number of neurons per layer, activation functions), and the ones that could be different for each client, mainly about optimization (optimizer, learning rate, batch size, number of epochs).

Because the first type of hyper-parameters must be common between all clients, they have to communicate some validation results in order to agree on their selection. For simplicity, the other type of hyper-parameters is also made common to all clients.

To that end, the \textit{collaborative grid search} is defined as a grid search in which the federated clients share their validation results for each considered set of hyper-parameters, so that the selected hyper-parameters are those that give the best results on average. Note that for the unsupervised solution, the model is validated only with benign data, so the selected hyper-parameters are those that minimize the loss. In the supervised solution, however, the selection is based on validation accuracy.

\subsubsection{Collaborative Threshold Selection}

For the unsupervised anomaly-detection approach, additionally to training the model, the clients have to select the threshold. To that end, in our proposed federated framework, each client $k$ computes a local threshold \change{on}{with} $\dkthr$ \addtxt{(using equation \ref{eq_threshold_mean_std})} and transmits it to the server, which then computes the global threshold as the average of the local thresholds. \addtxt{The global threshold is then given back to every client, which will use it for anomaly detection.} Note that this is not equivalent to computing the global threshold directly with the combination of all threshold-selection sets, as the threshold formula (\ref{eq_threshold_mean_std}) is non-linear (it uses the standard deviation). An alternative way of computing the threshold would be to share the whole set of MSE values over $\dkthr$ for each client $k$, and let the server compute the global threshold with that.

The threshold only needs to be computed for the final testing after the model has trained for the specified number of iterations. We still decided to compute it at several steps during the training in order to show its evolution.

\section{Adversarial Attacks and Countermeasures}
\label{sec:attacks}

This section provides the theoretical background regarding some of the most well-known poisoning attacks, intending to reduce the model performance. Besides, it also describes different model aggregation functions that could improve the resilience of the federated model training against attacks.

\subsection{Adversarial Attacks}

An honest server, a majority of honest clients and a minority of potentially colluding malicious clients are assumed through the following explanation.

Such malicious clients are often referred to as Byzantine workers \cite{byzantine}. The server and the honest participants could be considered as honest-but-curious as well (trying to infer as much information as possible without deviating from the protocol), but privacy issues are out of the scope of this work. Next, several data poisoning and model poisoning attacks are described, to be later implemented and evaluated in Section \ref{sec:experiments}. The characteristics of the described attacks are summarized in \tablename~\ref{tab:attacks}. \addtxt{These attacks have been selected for their simplicity and variety, but other more sophisticated and stealthy attacks exist in the literature \cite{adversarial_fl_lens, adversarial_fl_survey}.}

\begin{table}[h!]
\begin{center}
\setcellgapes{2pt}
\makegapedcells
\begin{tabular}{lccc}
\thline
\textbf{Attack name}  & \textbf{Poisoning} & \textbf{\makecell{Need\\data}} & \textbf{\makecell[c]{Attacker's\\objective}}           \\ \mhline
Benign label flipping & Data                 & Yes                & TNR = 0                        \\ \hline
Attack label flipping & Data                 & Yes                & TPR = 0                        \\ \hline
All labels flipping   & Data                 & Yes                & Acc. = 0                  \\ \hline
Gradient factor       & Model            & Yes                & Acc. = 0                        \\ \hline
Model cancelling       & Model            & No                 & \makecell[c]{$\wi = 0,$\\$\forall i \in [d]$}   \\ \hline \hline
\end{tabular}
\caption{Adversarial attack characteristics. The metrics shown here are defined in Section \ref{sec:experiments}.}
\label{tab:attacks}
\end{center}
\end{table}

Data poisoning attacks operate through the medium of the client dataset. The client could be malicious and intentionally modify its own data with the goal of making it misleading. Even if the client is honest, the attack could come from any part in the client data pipeline on which an external malicious entity has control. Therefore this attack category is the one that assumes the less from the clients and that is the most likely to happen. Three data poisoning attacks, all based on label flipping \cite{data_poisoning}, are described for the supervised situation. 
\begin{itemize}
    \item Benign label flipping. Here, the labels 0 (benign) are flipped to be 1s (attack). The goal of an attacker doing this would be to make the model always classify the traffic as attack and to make it have a TNR of 0\%. Such a model would constantly raise false alarms and could be very disturbing for its users.
    \item Attack label flipping. In this case, the labels 1 are flipped to be 0s, with the goal of making the model reach 0\% TPR. Such a model would never raise alarms about attack traffic and would allow potential malware to remain undetected.
    \item All labels flipping. In this attack all labels are flipped, \ie\ 1s become 0s and 0s become 1s. The goal of such an attack would be to bring the model accuracy to 0\%, combining both previous attacks.
\end{itemize}

Note that the two first attacks are considered as targeted since they focus on a specific class, while the third one is considered untargeted because it concentrates on both classes. All of these attacks are parameterized by the proportion $p_{poison}$ of the targeted labels that is flipped.

Model poisoning attacks are conducted through corrupted model updates sent to the server. They are a very big issue when using FL because the clients can send arbitrarily bad models to the server, and, due to the privacy that FL gives, it becomes hard to check whether the models received actually correspond to the local training data or not. In a sense, data poisoning attacks could be considered as a subset of model poisoning attacks because training from wrong data produces a wrong model. Next, some of the most basic model poisoning attacks are described:

\newcommand{\agrad}{\alpha_{grad}}
\newcommand{\aparam}{\alpha_{param}}

\begin{itemize}
    \item Gradient factor attack. In this case, the malicious clients multiply their gradients by a negative factor $\agrad$ before updating their local model and sharing it with the server. This attack is inspired by the omniscient attack in \cite{krum}, but instead of being aware of the estimate of the gradient, the malicious clients simply have access to the data from one device. Therefore, they are only able to compute the estimate of the gradient on their own data distribution. With $K$ total clients among which $f$ are malicious, the factor $\agrad$ is chosen to verify
    \begin{equation}
    \frac{1}{K}(K - f + \agrad \cdot f) = -1
    \label{eq_factor}
    \end{equation}
    Specifically, the malicious clients select their update factor so that the average update factor including honest clients is $-1$ (instead of $1$ in the non-adversarial case). Note that selecting a value of $\agrad$ that solves equation \ref{eq_factor} is not necessary in order to conduct this attack (any negative value could be considered).
    
    \item Model cancelling attack. In this attack, malicious clients try to bring all the global model parameters to the value $0$. They select their model in such a way that when averaged with the honest clients models, the original global model vanishes, i.e. $\wi = 0, \forall i \in [d]$. Only the most recent update from the honest clients remains. To that end, they simply output the original global model parameters, multiplied by a factor $\aparam$ that has to satisfy
    \begin{equation}
    K - f + \aparam \cdot f = 0
    \label{eq_param}
    \end{equation}
    Specifically, $\aparam$ must be selected so that the weight of the malicious clients ($\aparam \cdot f$) cancels the weight of the honest clients ($K - f$). Note that this time, using the right value of $\aparam$ is much more important, so collusion between the malicious clients is necessary so that they know their exact number ($f$) at the beginning. This attack is very powerful, but at the same time, it is not stealthy at all, as the values given by the malicious clients are very different from those usually expected in terms of direction and magnitude.
    
\end{itemize}

\subsection{Robust Model Aggregation Functions}
\label{subsection_robust_aggregation}
There are many different ways to make the system secure against attacks. One of the most extended ideas is to use model aggregation and update processing solutions that take into account the possibility of malicious clients trying to hijack the model \cite{adversarial_fl_survey}. Next, two different aggregation functions, in addition to averaging (AVG), are defined as well as a prior step to be applied to the models sent by the clients. Most of the convergence proofs of these aggregation functions do not hold in this work because the clients datasets are not from the same distributions. However, the intuition behind the use of these functions is still the same. \addtxt{These countermeasures have been selected for their great simplicity, as they only require a modification of the step of model aggregation, which is easy to implement. They also do not require any previous knowledge of the distribution of the client's data, which is hard to obtain in a realistic federated setting.}


\paragraph{Coordinate-wise median.} This aggregation function, as proposed by \cite{cm_tm}, applies the median to each parameter individually to exclude completely any potential outlier. The $i^{th}$ coordinate of $w$ is given by $\wi = med\{\wki : k \in [K]\}$. Note that the usual definition of the median is used, \ie\ when K is odd, the middle value is selected, and when K is even, the average between the two middle values is taken. We refer to this aggregation function as MED.

\paragraph{Coordinate-wise trimmed mean.} The trimmed mean, as proposed by \cite{cm_tm}, can be seen as a compromise between the averaging and the median. For each coordinate $i \in [d]$, a fraction of the largest and smallest values are removed before the mean is computed. Because of the low number of clients in the scenario that we consider, the trimmed mean algorithm is redefined using an integer number $c$ of excluded largest and lowest values instead of a proportion, but both are equivalent. Therefore, in our definition the $i^{th}$ coordinate of $w$ is given by $\wi = \frac{1}{K-2c}\sum_{u \in U^{(i)}}u$, where $U^{(i)}$ is a subset of $\{\wki : k \in [K]\}$ obtained by
removing the $c$ largest and the $c$ smallest of its elements. The number of excluded elements is $2c$. This aggregation function is referred as TM($c$).

\paragraph{s-Resampling.} Rather than being an aggregation function, s-Resampling \cite{s-resampling} is an additional step that can be done prior to the aggregation. In a scenario where each client's dataset has its own distribution, it aims at reducing the heterogeneity of the models sent by each client. Thus, s-Resampling is meant to be combined with a robust aggregation function to reduce the side-effects of using such a function on non-IID models. Note that combining s-Resampling with AVG is useless, as the result is always exactly the same as when only using AVG. It operates by replacing each model by the average between $s$ models randomly sampled from the $K$ clients models. Each model can be sampled a maximum of $s$ times in total. Algorithm \ref{algorithm_s_resampling} is a slightly adapted version of the second algorithm from \cite{s-resampling}. 
\begin{algorithm}[ht!]
\SetAlgoLined
\SetAlgoNoEnd
\SetKwInOut{Input}{input}
\Input{$\{\wk : k \in [K]\}$, $s$, $\{c[k] \coloneqq 0 : k \in [K]\}$}
  \For{$k^{\prime} \coloneqq 1, \dots, K$}{
    \For{$i \coloneqq 1, \dots, s$}{
      \While{Select $j_{i} \sim $  \textbf{Uniform}($[K]$)}{
        \If{$c[j_{i}] < s$}{
          $c[j_{i}] \mathrel{+}= 1$\\
          break}
        }
  }
  Compute average $\bar{w}_{k^{\prime}} \coloneqq \frac{1}{s}\sum_{i=1}^{s}w_{j_{i}}$
 }
 \Return $\{\bar{w}_{k^{\prime}} : k^{\prime} \in [K]\}$
 \caption{Resampling with s-replacement\\source: \cite{s-resampling}}
 \label{algorithm_s_resampling}
\end{algorithm}
Indeed, s-Resampling may also cause the malicious models to be diluted into several of the models that it outputs, increasing the reach of the malicious clients. For that reason, it is only expected to work satisfyingly with a small number of malicious clients, a small value of $s$, and with an aggregation function that gets rid of a high number of extreme values, such as MED or TM($2$).

\section{Experimental Results}


\label{sec:experiments}

This section details the results obtained in the different experiments performed to validate the proposed framework. First, it compares the performance when detecting malware between federated and traditional approaches when following both a supervised or an unsupervised solution. After that, it shows the impact of the adversarial attacks proposed in Section \ref{sec:attacks}, and how the different aggregation functions mitigate those attacks.

The metrics used to evaluate and compare the performance of each approach are the following (\textit{TP}: True Positives, \textit{TN}: True Negatives, \textit{FP}: False Positives, \textit{FN}: False Negatives, \textit{TPR}: True Positive Rate, \textit{TNR}: True Negative Rate):

\newcommand\sbullet[1][.5]{\mathbin{\vcenter{\hbox{\scalebox{#1}{$\bullet$}}}}}
\begin{itemize}
    \item $\textit{TPR}=\frac{\textit{TP}}{\textit{TP}+\textit{FN}}$ \hspace{0.5cm} $\sbullet[1.38]$ \hspace{0.02cm} $\textit{Accuracy}=\frac{\textit{TP}+\textit{TN}}{\textit{TP}+\textit{FP}+\textit{TN}+\textit{FN}}$
    \item $\textit{TNR}=\frac{\textit{TN}}{\textit{TN}+\textit{FP}}$ \hspace{0.45cm} $\sbullet[1.38]$ \hspace{0.02cm} $\textit{F1-Score}=\frac{\textit{TP}}{\textit{TP}+\frac{1}{2}(\textit{FP}+\textit{FN})}$
\end{itemize}

All experiments performed in this section have followed a similar methodology. In this sense, the federation consists of $K=8$ clients, each owning data of one of the 9 devices available in the N-BaIoT dataset. The data of one device was not used during training, keeping it as an unseen device for testing purposes. In this context, nine different combinations of devices (with an unseen one) were used in all experiments. Moreover, experiments were repeated 5 times to improve the consistency of results. Finally, the results of each experiment show the average over 45 runs in total (9 possible unseen devices and 5 executions).

\subsection{Performance of Federated and Traditional Learning for the Detection of Malware in IoT Devices}
\label{sec:exp_performance}

\newcommand{\decnaive}{\makecell[c]{Naive}}
\newcommand{\decminib}{\makecell[c]{\textsc{Mini-}\\\textsc{batch}\\\textsc{avg}}}
\newcommand{\decmultie}{\makecell[c]{\textsc{Multi-}\\\textsc{epoch}\\\textsc{avg}}}
\newcommand{\centm}{\makecell[c]{Central.}}
\newcommand{\knowndev}{\makecell[c]{Known\\devices}}
\newcommand{\newdev}{\makecell[c]{New\\device}}
\newcommand{\accuracy}{Acc.}
\newcommand{\valsuf}{}

This experiment seeks to measure the performance of our solution when detecting IoT malware using N-BaIoT dataset. To verify that the federated learning approach fits our IoT malware scenario properly, it is necessary to compare it with traditional solutions. More specifically, the compared alternatives are:

\begin{itemize}
    \item Naive decentralized approach. Each client uses its local dataset for training and testing. Since each client produces its own model, the results are compared by averaging the performance of each client.
    \item Centralized approach. All training data is shared with a server in charge of training and testing a model with it. It does not preserve privacy.
    \item Federated with \minibavg. The different clients collaborate to generate a global model using the \minibagg\ algorithm with the AVG aggregation function.
    \item Federated with \multieavg. Similar to the previous approach but training with the \multieagg\ algorithm in order to greatly reduce the communication costs.
\end{itemize}

The following steps are followed both for the supervised and the unsupervised solutions. First, two important hyper\addtxt{-}parameters (the architecture of the model and the L2- regularization value $\lambda$) are selected for each setup using grid searches. The MLP and autoencoder architectures considered are those described in Section~\ref{subsubsection_model_training}. The values considered for $\lambda$ are $0$, $10^{-5}$ and $10^{-4}$. Note that for the naive method, the hyper-parameters are selected per client because the clients do not collaborate on hyper-parameter selection. For the FL approaches, each federation used collaborative grid searches to select the hyper-parameters, as defined in Section \ref{subsubsection_collaborative_gs}. Finally, in the centralized method, the grid search is performed directly by the server that receives the whole dataset. For all experiments, a batch size of $B=64$ was used when training, except with \minibavg\ where the batch size was divided by the number of clients ($B=8$), so that each model update is made with a total of $64$ samples as well. In all of the experiments, the model updates are computed with Stochastic Gradient Descent (SGD). For the supervised solution, the training is conducted for $E=4$ epochs; for the unsupervised solution, it is made with $E=120$ epochs.

\paragraph{Supervised situation.}
First, the supervised solution is verified. Here, the three different dataset splitting options explained in Section~\ref{sec:dataset_splitting} (7.87\%, 50\% and 95\% benign data) are used in repeated tests, also checking how the different class balances affect the results.
\tablename~\ref{tab:supervised_averaging} shows the results achieved in these experiments.

\aboverulesep=-0.75px
\belowrulesep=-0.75px

\begin{table}[hb!]
\begin{center}
\setcellgapes{\cellgps}
\makegapedcells
\begin{tabular}{cl|cccc^}
\cmidrule[1.8\arrayrulewidth]{3-6}
\multicolumn{1}{l}{} & \multicolumn{1}{c^}{} & \multicolumn{1}{c}{\decnaive}     & \decmultie & \decminib & \multicolumn{1}{c^}{\centm} \\ \cmidrule[1.8\arrayrulewidth]{3-6} \noalign{\vskip 2mm} \mhline
\multicolumn{1}{^c|}{\multirow{3}{*}{\makecell[c]{Known\\devices\\\textcolor{red}{$(7.87\%)$}}}} & \accuracy & 99.78\valsuf & 99.92\valsuf & 99.96\valsuf & 99.96\valsuf \\ \cline{2-6} 
\multicolumn{1}{^c|}{}                               & TPR               & 99.98\valsuf & 99.98\valsuf & 99.98\valsuf & 99.98\valsuf \\ \cline{2-6} 
\multicolumn{1}{^c|}{}                               & TNR               & 97.49\valsuf & 99.26\valsuf & 99.69\valsuf & 99.69\valsuf \\ \mhline
\multicolumn{1}{^c|}{\multirow{3}{*}{\makecell[c]{New\\device\\\textcolor{red}{$(7.87\%)$}}}}    & \accuracy & 98.94\valsuf & 99.89\valsuf & 99.89\valsuf & 99.88\valsuf \\ \cline{2-6} 
\multicolumn{1}{^c|}{}                               & TPR               & 99.58\valsuf & 99.97\valsuf & 99.98\valsuf & 99.97\valsuf \\ \cline{2-6} 
\multicolumn{1}{^c|}{}                               & TNR               & 91.44\valsuf & 99.00\valsuf & 98.93\valsuf & 98.83\valsuf \\ \mhline \noalign{\vskip 1.5mm} \mhline
\multicolumn{1}{^c|}{\multirow{3}{*}{\makecell[c]{Known\\devices\\\textcolor{ForestGreen}{$(50\%)$}}}} & \accuracy & 99.92\valsuf & 99.82\valsuf & 99.93\valsuf & 99.91\valsuf \\ \cline{2-6} 
\multicolumn{1}{^c|}{}                               & TPR               & 99.97\valsuf & 99.97\valsuf & 99.97\valsuf & 99.97\valsuf \\ \cline{2-6} 
\multicolumn{1}{^c|}{}                               & TNR               & 99.88\valsuf & 99.67\valsuf & 99.88\valsuf & 99.85\valsuf \\ \mhline
\multicolumn{1}{^c|}{\multirow{3}{*}{\makecell[c]{New\\device\\\textcolor{ForestGreen}{$(50\%)$}}}}    & \accuracy & 98.36\valsuf & 99.63\valsuf & 99.58\valsuf & 99.52\valsuf \\ \cline{2-6} 
\multicolumn{1}{^c|}{}                               & TPR               & 98.79\valsuf & 99.90\valsuf & 99.95\valsuf & 99.93\valsuf \\ \cline{2-6} 
\multicolumn{1}{^c|}{}                               & TNR               & 97.93\valsuf & 99.35\valsuf & 99.21\valsuf & 99.10\valsuf \\ \mhline \noalign{\vskip 1.5mm} \mhline
\multicolumn{1}{^c|}{\multirow{3}{*}{\makecell[c]{Known\\devices\\\textcolor{blue}{$(95\%)$}}}} & \accuracy & 99.92\valsuf & 99.79\valsuf & 99.93\valsuf & 99.93\valsuf \\ \cline{2-6} 
\multicolumn{1}{^c|}{}                               & TPR               & 99.89\valsuf & 99.93\valsuf & 99.98\valsuf & 99.98\valsuf \\ \cline{2-6} 
\multicolumn{1}{^c|}{}                               & TNR               & 99.92\valsuf & 99.78\valsuf & 99.92\valsuf & 99.93\valsuf \\ \mhline
\multicolumn{1}{^c|}{\multirow{3}{*}{\makecell[c]{New\\device\\\textcolor{blue}{$(95\%)$}}}}    & \accuracy & 98.59\valsuf & 99.43\valsuf & 99.38\valsuf & 99.42\valsuf \\ \cline{2-6} 
\multicolumn{1}{^c|}{}                               & TPR               & 97.79\valsuf & 99.55\valsuf & 99.82\valsuf & 99.81\valsuf \\ \cline{2-6} 
\multicolumn{1}{^c|}{}                               & TNR               & 98.63\valsuf & 99.42\valsuf & 99.36\valsuf & 99.40\valsuf \\ \mhline
\end{tabular}
\end{center}
\caption{Supervised results comparing both FL approaches (\textsc{Multi-epoch avg} and \textsc{Mini-batch avg}) with the naive approach and the centralized approach. The percentages of benign data of the datasets used are indicated in color on the left.}
\label{tab:supervised_averaging}
\end{table}

\begin{figure}[ht!]
\centering
\begin{subfigure}{.44\textwidth}
  \centering
  \includegraphics[width=.99\linewidth]{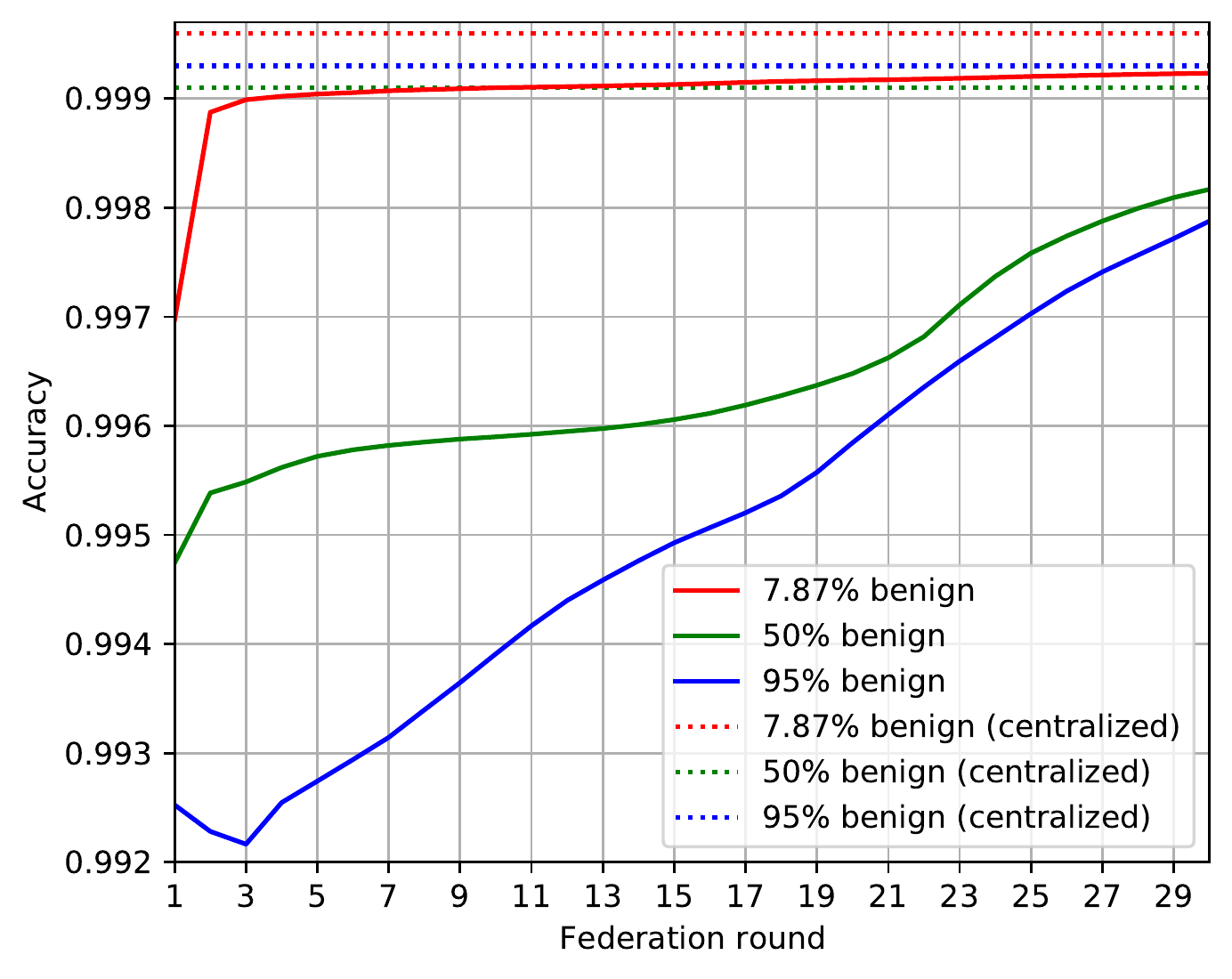}
  \caption{Over the federation round (\multieavg)}
\end{subfigure}

\begin{subfigure}{.44\textwidth}
  \centering
  \includegraphics[width=.99\linewidth]{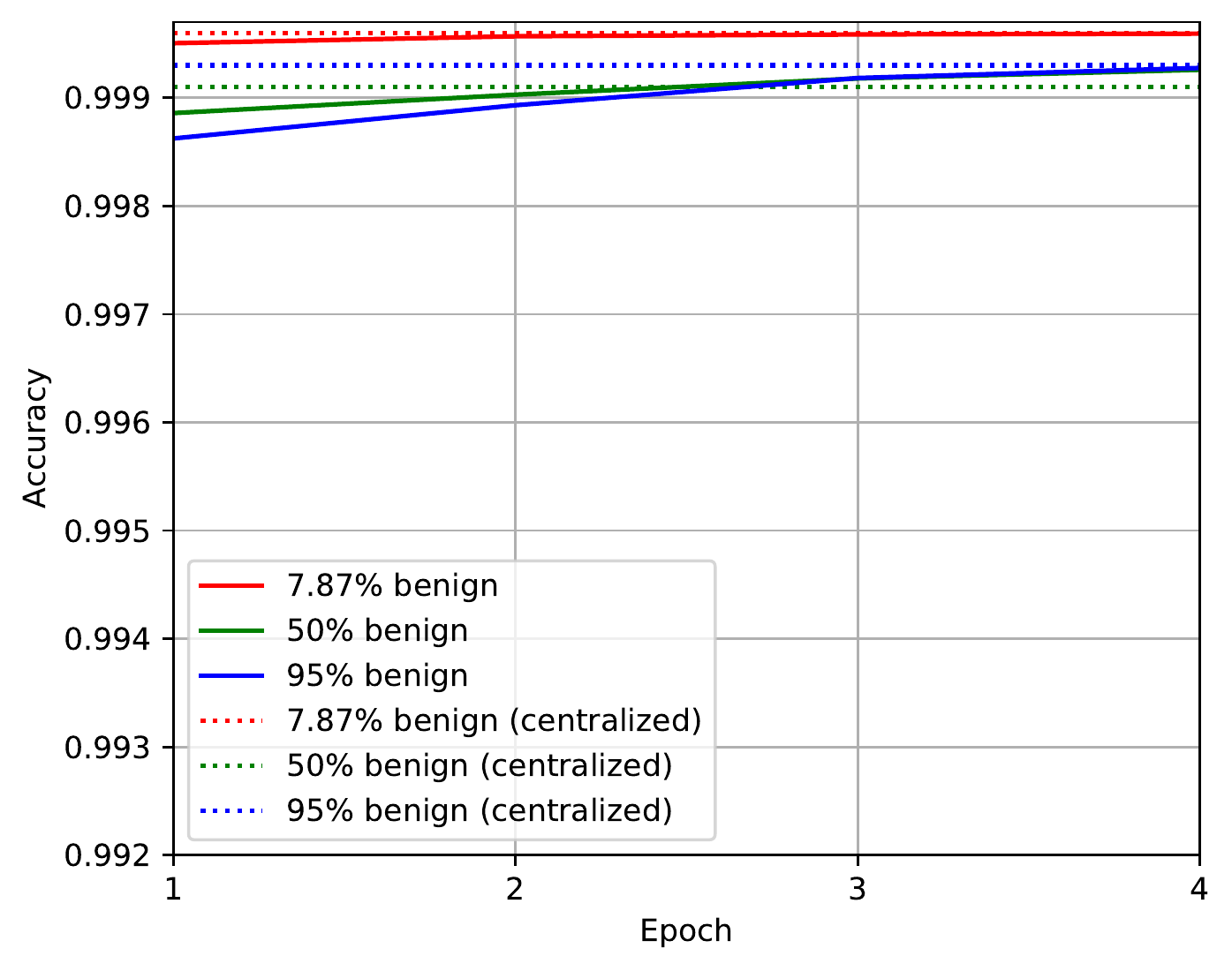}
  \caption{Over the training epoch (\minibavg)}
\end{subfigure}
\caption{Evolution of the known devices accuracy. The accuracies obtained with the centralized methods are displayed with dotted lines for comparison.}
\label{fig:supervised_accuracy_fl}
\end{figure}

The first noticeable result is that the centralized method's performance is higher than the distributed naive one, especially when evaluated on an unseen device. Moreover, on all three dataset settings, the \minibavg\ results are very close to the centralized ones, even sometimes exceeding them. Although obtaining better results than in the centralized method could be surprising, this can be explained by several factors, such as the randomness of the experiments or the fact that the hyper-parameters are computed differently. \figurename~\ref{fig:supervised_accuracy_fl} shows how fast the models converge near the centralized performance.

\multieavg\ also produces quite satisfying results, with an insignificant decrease in the accuracy on known devices, compensated by an accuracy always exceeding the centralized one on the new device. This can be explained by the fact that averaging the model parameters with a non-convex loss function can have arbitrarily damaging effects on the model, as explained in \cite{fedavg}. However, it can also be viewed as a form of mechanism acting against overfitting, thus improving the generalization on a new device. Moreover, these results could probably be slightly improved by using a higher number $T$ of federation rounds. \figurename~\ref{fig:supervised_accuracy_fl} shows that for the datasets with 50\% and 95\% benign data, the accuracy seems to have not exactly converged after 30 rounds.

\paragraph{Unsupervised situation.}
\hspace{-0.05cm}Once the supervised performance 
is verified, the next step is to evaluate the unsupervised one. 
Here, only the benign traffic is used for training, so the final model does not depend on the class balance in the dataset. In order to make our results independent of the class balance used, we only show the TPR and the TNR for this solution (and not the accuracy).
The equation used to define the threshold is described in Section \ref{subsubsection_model_training}.
Among the possible architectures, the first one (Autoencoder A) is always the one giving the best validation loss during all hyper-parameter selections
. All results from the unsupervised situation are further produced with Autoencoder A. \tablename~\ref{tab:unsupervised_averaging_mean+std} shows the unsupervised results of the system.

\begin{table}[h!]
\begin{center}
\setcellgapes{\cellgps}
\makegapedcells
\begin{tabular}{cl|cccc^}
\cmidrule[1.8\arrayrulewidth]{3-6}
\multicolumn{1}{l}{} & \multicolumn{1}{c^}{} & \multicolumn{1}{c}{\decnaive}     & \decmultie & \decminib & \multicolumn{1}{c^}{\centm} \\ \cmidrule[1.8\arrayrulewidth]{3-6} \noalign{\vskip 1.5mm} \mhline
\multicolumn{1}{^c|}{\multirow{2}{*}{\knowndev}} & TPR               & 88.00\valsuf & 99.98\valsuf & 99.98\valsuf & 99.98\valsuf \\ \cline{2-6} 
\multicolumn{1}{^c|}{}                               & TNR               & 97.38\valsuf & 94.84\valsuf & 95.12\valsuf & 95.56\valsuf \\ \mhline
\multicolumn{1}{^c|}{\multirow{2}{*}{\newdev}}    & TPR               & 87.77\valsuf & 99.98\valsuf & 99.98\valsuf & 99.98\valsuf \\ \cline{2-6} 
\multicolumn{1}{^c|}{}                               & TNR               & 59.66\valsuf & 92.61\valsuf & 91.78\valsuf & 92.76\valsuf \\ \mhline
\end{tabular}
\end{center}
\caption{Unsupervised results comparing both FL approaches (\textsc{Multi-epoch avg} and \textsc{Mini-batch avg}) with the naive approach and the centralized approach.}
\label{tab:unsupervised_averaging_mean+std}
\end{table}

\begin{figure}[ht!]
\centering
\begin{subfigure}{.43\textwidth}
  \centering
  \includegraphics[width=.99\linewidth]{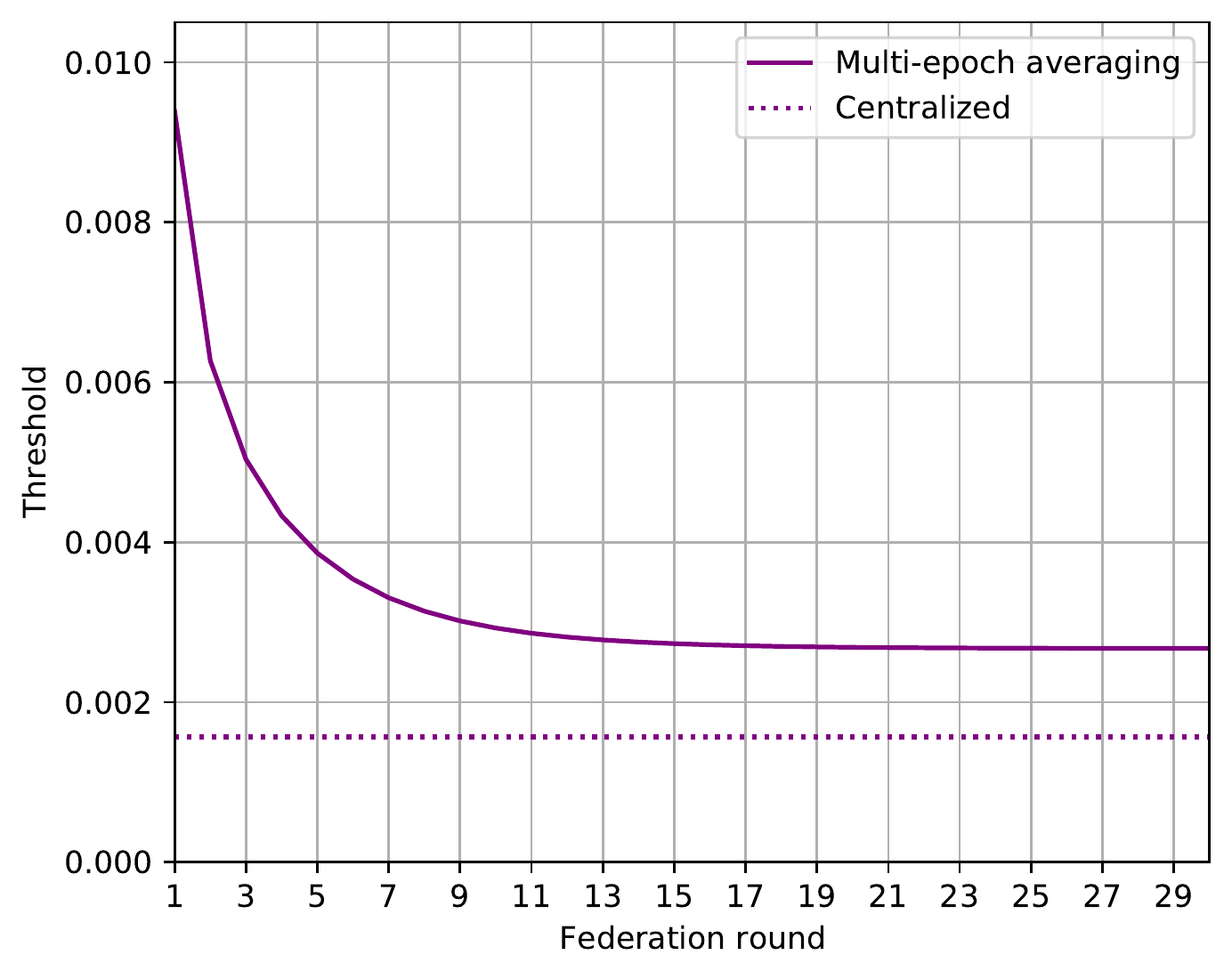}
  \caption{Over the federation round (\multieavg)}
\end{subfigure}

\begin{subfigure}{.43\textwidth}
  \centering
  \includegraphics[width=.99\linewidth]{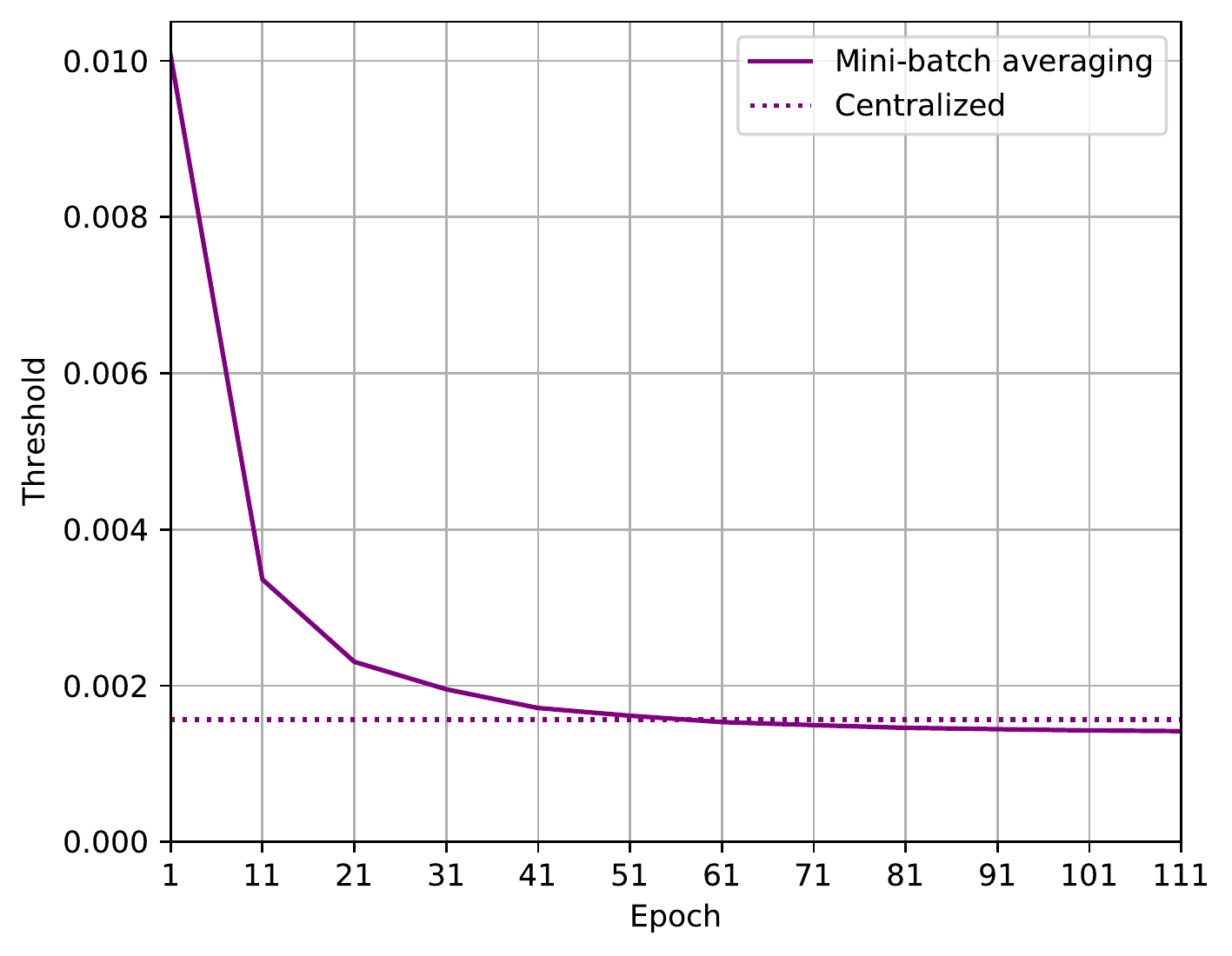}
  \caption{Over the training epoch (\minibavg)}
\end{subfigure}
\caption{Evolution of the global threshold values with both FL algorithms. The threshold obtained in the centralized method is displayed with a dotted line for comparison.}
\label{fig:thresholds}
\end{figure}

Here centralizing the data presents overall a high performance improvement over the naive method. Furthermore, the FL algorithms also very successfully deal with the unsupervised fingerprinting task. Specifically, the centralized performance is almost reached by both the \multieavg\ and the \minibavg\ methods. Once again, \multieavg\ seems to help the model to generalize better, as it demonstrates on the new device a marginally better TNR than \minibavg. Interestingly, the threshold, as displayed in \figurename~\ref{fig:thresholds}, converges to a larger value, in the case of \multieavg\, than what the centralized method achieves. As explained earlier, the collaborative threshold selection is not equivalent to selecting the threshold directly on the whole threshold-selection set (as in the centralized method), so this result is not surprising.

With the previous experiments, it has been verified that in this particular scenario of malware detection in IoT devices, using more data to train the model presents a significant improvement, especially on previously unseen devices. Besides, FL-based training successfully reaches the centralized performance in a privacy-preserving manner.

\subsection{Impact of Adversarial Attacks and Countermeasures when Detecting Malware}

Once the performance of the federated approach has been verified, the next step is to evaluate how the different adversarial attacks proposed in Section \ref{sec:attacks} affect the federated approach. Besides, different aggregation functions are applied to test how they improve the model resilience against the different attacks.
For conciseness, these experiments focus on the supervised situation and use only the dataset balance with 95\% of benign data. Moreover, they are conducted with the \minibagg\ federated algorithm. A batch size of $B=64$ (instead of $B=8$) is used for all the adversarial experiments, as it allows smoother updates for the robust aggregation functions.

As explained earlier, s-Resampling is only expected to work with a small value of $s$, and combined with MED or TM($2$) (or other robust aggregation functions that we did not implement). Because TM($2$) computes its output by taking more values into account than MED, s-Resampling was only experimented for TM($2$) and with $s=2$. This combination is referred as TM($2$) $\circ$ 2-Resampling.

In the experiments implementing data poisoning attacks, the \textit{All labels flipping} attack is selected for testing, as it combines both benign and attack label flipping. Since the focus is placed on intentional data poisoning, $p_{poison} = 1$ is always used. This approach enables the verification of the maximum impact of the attack in the generated model.

Regarding model poisoning attacks, in the case of \textit{gradient factor} attack, solving equation \ref{eq_factor} gives $\agrad = \frac{f - 2K}{f}$. For a total of $8$ clients including $1$, $2$, and $3$ malicious clients, the values chosen for $\agrad$ are respectively $-15$, $-7$ and $-\frac{13}{3}$. In the case of \textit{model cancelling} attack, solving equation \ref{eq_param} gives $\aparam = \frac{f-K}{f}$. For a total of $8$ clients including $1$, $2$, and $3$ malicious clients, the values chosen for $\aparam$ are respectively $-7$, $-3$ and $-\frac{5}{3}$.

\begin{figure}[htpb!]
\centering
\begin{subfigure}{.48\textwidth}
\centering
\includegraphics[width=.99\linewidth]{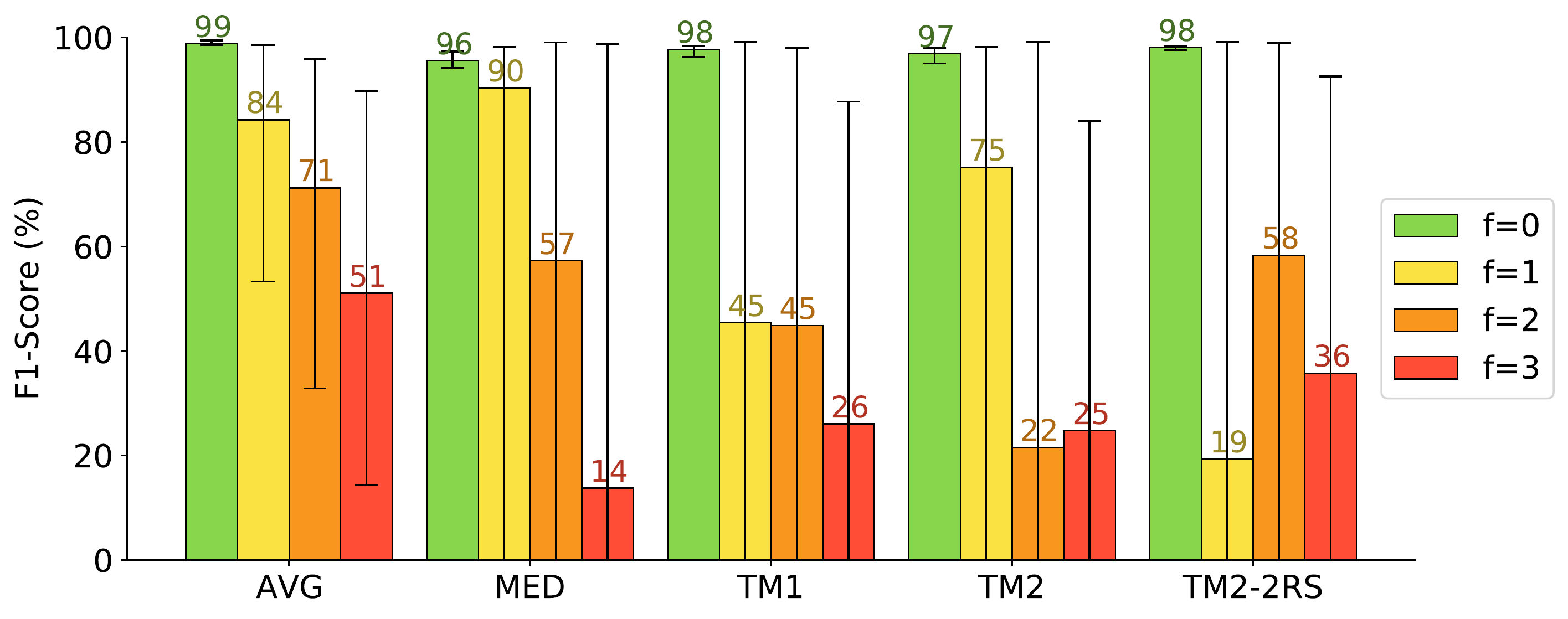}
\caption{All labels flipping attack.}
\label{fig:all_label_flipping_f1}
\end{subfigure}

\begin{subfigure}{.48\textwidth}
\centering
\includegraphics[width=.99\linewidth]{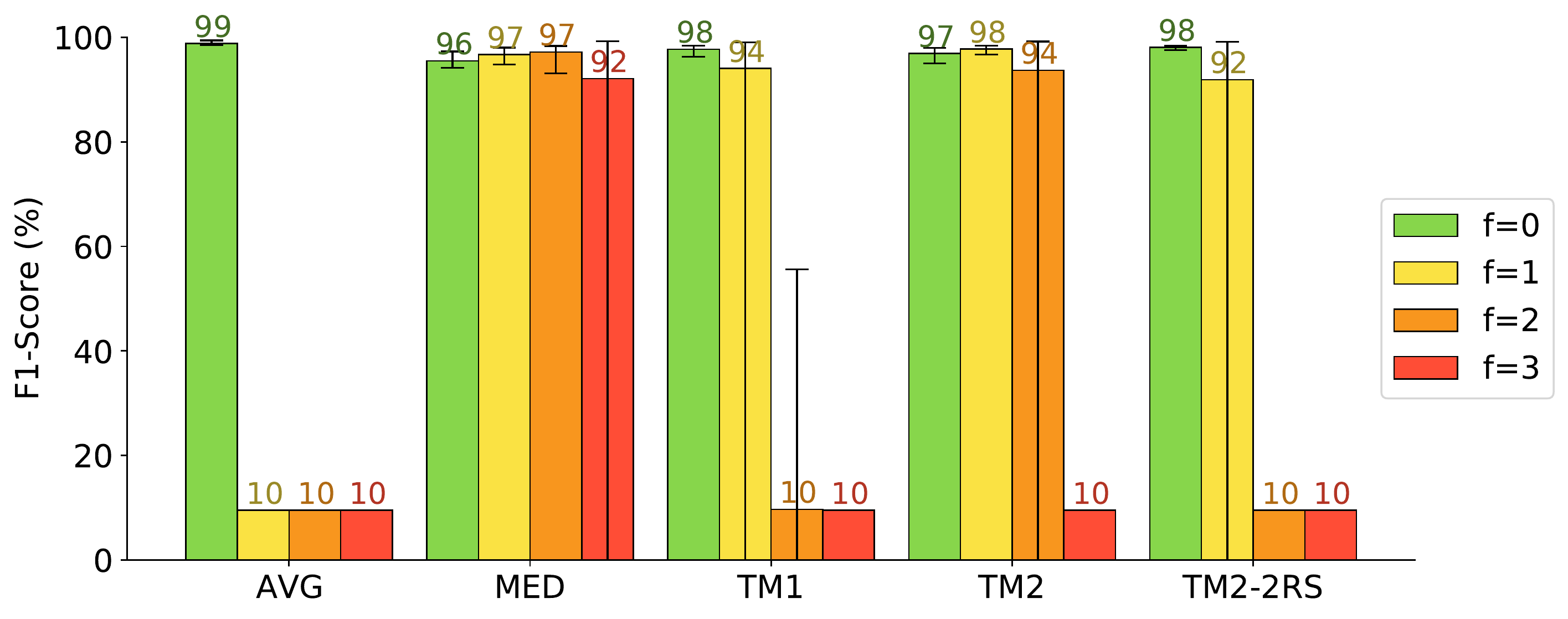}
\caption{Gradient factor attack.}
\label{fig:gradient_factor_f1}
\end{subfigure}

\begin{subfigure}{.48\textwidth}
\centering
\includegraphics[width=.99\linewidth]{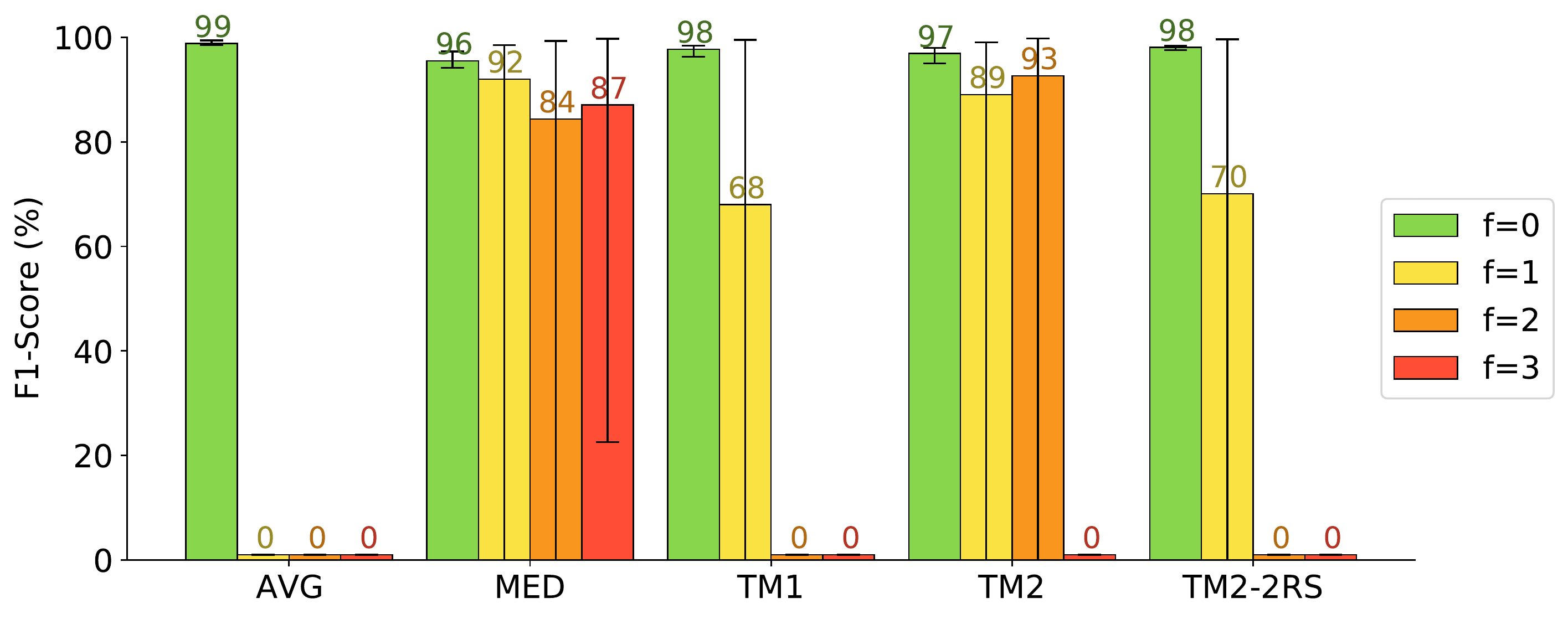}
\caption{Model cancelling attack.}
\label{fig:model_cancelling_f1}
\end{subfigure}
\caption{F1-Scores under the different tested attacks for each aggregation function, with $f = 0$, $1$, $2$ or $3$ malicious clients (respectively $0\%$, $12.5\%$, $25\%$ and $37.5\%$ of the total clients). 
The minimum and maximum values (over the 45 runs) of the F1-Scores are displayed with capped bars.}
\label{fig:attack_f1}
\end{figure}

\figurename~\ref{fig:attack_f1} shows how the F1-Score of the model tested on the devices owned by the clients (the known devices) varies in the different implemented attacks according to the aggregation function. It also shows the evolution of the metric when the number of malicious clients grows from 0 to 3 (or 37.5\% of the total number of clients in the setup). It has to be noted that these results have a large variance due to the randomness of the selection of which client is malicious. Even though each experiment was run a total of 45 times, this can lead to unanticipated results, such as sometimes having a better average F1-Score with more malicious clients. Nonetheless, these experiments are sufficient to get a well-founded idea of the seriousness of the adversarial problem.

As we can observe, \textit{averaging} (AVG) is the best aggregation function when all clients are honest. However, when malicious clients are involved, its performance is heavily affected depending on the attack. Specifically, under the \gradientf\ and the \modelc\ attacks, even a single malicious client is sufficient to consistently turn the model into a constant predictor (note that a constant positive predictor has an F1-Score of $\sim$10\% and a constant negative predictor has an F1-Score of 0\%). This demonstrates the necessity of using more robust methods when assuming a threat model in which even only one client could be malicious.

\textit{Coordinate-wise median} aggregation (MED) presents more resilience against most attack scenarios considered. Overall it has the best results among the tested aggregation functions in the adversarial setup. However, this is still far from being robust enough, especially when considering 3 malicious clients, as the \alllf\ attack makes its F1-Score reach an average value of around 14\%. 
Even with a single malicious client, when performing \alllf\ and \modelc\ attacks, although the average F1-Scores are respectively 90\% and 92\%, their minimum value over the 45 runs is 0\% in both cases, making it still highly unreliable.

Unsurprisingly, \textit{Coordinate-wise trimmed mean} (TM($c$)) fails when used against more than $c$ malicious clients, as clearly demonstrated in the \modelc\ attack results (\figurename~\ref{fig:model_cancelling_f1}). However, it does not mean that this aggregation function performs well when $c \geq f$, as the minimum F1-Score reaches 0\% even for a single malicious client under the \alllf\ attack (\figurename~\ref{fig:all_label_flipping_f1}). The only benefit of TM(1) and TM(2) over MED lies in the performance when no malicious client is involved, which is a bit better as more parameters are considered during the computation of the global model. This advantage might be higher in a use case with more clients, but in our case it is too low to justify the usage of TM.

Finally, \textit{2-Resampling} shows an improvement of accuracy on the known devices when no malicious client is involved. However, this comes at the cost of reducing the robustness of the system, making TM(2) $\circ$ 2-Resampling have similar results as TM(1) most of the time. Still, a small improvement over TM(2), shown in \figurename~\ref{fig:all_label_flipping_f1}, has to be noted with 2 and 3 malicious clients. Similarly to TM, s-Resampling does not provide enough advantage to be used in such a small federation, but it could become more useful at a larger scale.

As general remarks, although the resilience of the models has been greatly improved using MED under model poisoning attacks (\textit{gradient factor} and \textit{model cancelling} attack\addtxt{s}), the performance of the model is still reduced substantially. In addition, in the case of the \textit{all labels flipping} attack, AVG still performs better than other functions that seek to improve the robustness of the model. 
These results show that, although the model performance has been improved, further research on aggregation functions that are resilient to adversarial attacks is still required.
\addtxt{We believe that in the case of other attacks that greatly affect the weights, results would be similar to those of \textit{gradient factor} and \textit{model cancelling} attacks. It is however unknown how performance would be affected in the case of other more sophisticated and stealthy attacks.}

\section{Discussion}
\label{sec:discussion}

This section discusses relevant aspects of performance and architecture design that must be considered when deployed on a real B5G environment. Although the performance in the malware detection experiments has proven to be high, aspects such as communication costs or framework centralization should be discussed.

\subsection{Number of clients and adversarial results}

One of the limitations in the experimentation has been the low number of clients used, 8 for training, due to the availability of datasets suitable for federated learning. \addtxtRev{In a real B5G scenario, device deployments will reach up to 10M devices per km$^2$ according to ITU (International Telecommunication Union) requirements \cite{series2017minimum}.} However, we consider that the experiments are valid since, although the number of adversaries is low, namely 1, 2 and 3, the percentage they represent over the total number of clients performing the training is relatively high, 12.5\%, 25\% and 37.5\%, respectively (see \figurename~\ref{fig:attack_f1}).  Thus, the results can be extrapolated to environments with a much larger number of clients but where the adversaries represent a small percentage of the total, no more than 50\%.

\addtxt{In addition, other robust aggregation algorithms should be tested since the current ones do not offer sufficient attack resilience when the number of malicious clients exceeds 25\%. In this regard, there are interesting proposals on aggregation algorithms that already take into account the possible presence of malicious clients and evaluate variations in the models it shares. The most interesting ones to be assessed as future work are Krum \cite{krum}, Bulyan \cite{bulyan} and AUROR \cite{auror}.}

\addtxt{\subsection{Communication and computation costs}}

\addtxtRev{Although B5G throughput requirements (100 Mbps in \cite{series2017minimum}) exceed by far the requirements of the proposed solution,} since the framework is designed for clients to be located at or near the access points, communication and computation costs \addtxtRev{should be considered. They} are critical in order not to influence the regular operation of the wireless interfaces of the IoT objects and the network elements that provide access to them.

\addtxt{\minibagg\ has much higher communication costs than \multieagg\, as it requires $E \cdot \frac{n_{k}}{B}$ model transmissions per client for the full training, where $B$ is the batch size, $E$ is the number of epochs and $n_{k}$ is the number of training samples of client $k$. Note that in terms of computation cost, it also indicates the number of local model updates performed by each client. On the other hand, \multieagg\ just requires the clients to transmit the model to the server once per round, for a total of $T$ transmissions per client. However, the number of local model updates is also $T$ times larger, \ie\ $T \cdot E \cdot \frac{n_{k}}{B}$ for client $k$.}

\addtxt{\tablename~\ref{tab:comm} shows the comparison between both aggregation algorithms in the experiments of Section \ref{sec:exp_performance} in terms of computation and communication costs. \multieavg\ shows much lower communication costs than \minibavg, $\approx$1300 times less in the case of the supervised approach and $\approx$2000 times less in the unsupervised counterpart. However, the number of local training iterations is $3.75$ times higher for the hyper-parameters that have been selected. The throughput in a real 5G or B5G network should be sufficient to deploy any of the two algorithms. In addition, as stated in Section \ref{sec:architecture}, the framework clients will be B5G base stations and other access points, which have a relatively high computational power. However, if the communication cost becomes a critical issue, it would be natural to opt for an approach based on \multieavg.}


\begin{table}[htpb]
\raggedright
\begin{subtable}{0.95\columnwidth}
    \centering
    \begin{tabular}{l|c|c^}
    \cmidrule[1.8\arrayrulewidth]{2-3}
    \multicolumn{1}{c^}{} & \makecell[c]{\textsc{Multi-epoch}\\\textsc{avg}} & \multicolumn{1}{c^}{\makecell[c]{\textsc{Mini-batch}\\\textsc{avg}}} \\
    \cmidrule[1.8\arrayrulewidth]{2-3} \noalign{\vskip 1.5mm} \mhline
    \multicolumn{1}{^c|}{\makecell[c]{Number of model\\transmissions}} & $T = 30$ & \makecell[c]{$E \cdot \frac{n_{k}}{B}$\\$=4 \cdot 9875$\\$=39500$}\valsuf \\ \mhline
     \multicolumn{1}{^c|}{\makecell[c]{Communication cost \\ assuming a model\\ of size 94 kB}} & 2.82 MB & 3.713 GB\valsuf \\ \mhline
      \multicolumn{1}{^c|}{\makecell[c]{Number of local\\training steps}} & \makecell[c]{$T \cdot E \cdot \frac{n_{k}}{B}$ \\ $\simeq 30 \cdot 4 \cdot 1234 $ \\ $=148080$} & \makecell[c]{$E \cdot \frac{n_{k}}{B}$\\$=4 \cdot 9875$\\$=39500$}\valsuf \\ \mhline
    \end{tabular}
    \caption{\addtxt{Computation and communication costs in the supervised approach. When using \multieavg, $\frac{n_{k}}{B}=\frac{79000}{64} \simeq 1234$, and using \minibavg, $\frac{n_{k}}{B}=\frac{79000}{8} = 9875$.}\\}
    \end{subtable}
    \begin{subtable}{0.95\columnwidth}
    \centering
       \begin{tabular}{l|c|c^}
    \cmidrule[1.8\arrayrulewidth]{2-3}
    \multicolumn{1}{c^}{} & \makecell[c]{\textsc{Multi-epoch}\\\textsc{avg}} & \multicolumn{1}{c^}{\makecell[c]{\textsc{Mini-batch}\\\textsc{avg}}} \\
    \cmidrule[1.8\arrayrulewidth]{2-3} \noalign{\vskip 1.5mm} \mhline
    \multicolumn{1}{^c|}{\makecell[c]{Number of model\\transmissions}} & \makecell[c]{$T = 30$ } & \makecell[c]{$E \cdot \frac{n_{k}}{B}$ \\ $\simeq 120 \cdot 494 $ \\ $=59280$} \valsuf \\ \mhline
     \multicolumn{1}{^c|}{\makecell[c]{Communication cost \\ assuming a model\\ of size 27 kB}} & 810 kB & 1.6 GB\valsuf \\ \mhline
      \multicolumn{1}{^c|}{\makecell[c]{Number of local\\training steps}} & \makecell[c]{$T \cdot E \cdot \frac{n_{k}}{B}$ \\ $\simeq 30 \cdot 120 \cdot 62 $ \\ $=223200$} & \makecell[c]{$E \cdot \frac{n_{k}}{B}$ \\ $\simeq 120 \cdot 494 $ \\ $=59280$}\valsuf \\ \mhline
    \end{tabular}
            \caption{\addtxt{Computation and communication costs in the unsupervised approach. When using \multieavg, $\frac{n_{k}}{B}=\frac{3950}{64} \simeq 62$, and using \minibavg, $\frac{n_{k}}{B}=\frac{3950}{8} \simeq 494 $.}}
    \end{subtable}%
    \caption{\addtxt{Computation and communication costs per client. The communication cost is from the client's perspective and has to be considered in both directions (download and upload). The assumed model sizes correspond to the largest architectures that were used in our experiments, for both the supervised and the unsupervised approaches.}}
    \label{tab:comm}
\end{table}

\addtxt{\subsection{Decentralization and non-synchronization}}

Although the training of the models is decentralized at each client, having a server in charge of model aggregation has many advantages, such as controlling the common model generated, coordination between clients, etc. However, this design also brings with it some disadvantages.

\addtxt{The server becomes a central point of failure, where a bottleneck or attack can make it no longer possible to aggregate the local models, and only the local models can be used on each client. Therefore, it is necessary to ensure the correct scaling of the server functionalities to ensure that there are no bottlenecks and use the appropriate security solutions to prevent attacks on the server as much as possible. An additional solution would be to adapt the platform towards a purely decentralized approach where models are shared using Blockchain and each client performs the aggregation locally, eliminating the need for a coordinator in the process.}

\addtxt{Another disadvantage is the synchronization required between clients when submitting their models for aggregation. A client that fails or is slow due to asynchrony may cause the server not to perform the training correctly \cite{al2020iot}. Currently, the framework addresses this problem by setting a timeout for sending the models so that if one of the clients does not respond in time, it is skipped from that aggregation step. In this case, a Blockchain-based solution is also beneficial since it can be used as an asynchronous repository where each client can publish its models each time it trains them locally.}

\addtxt{Despite its benefits to solve both disadvantages, it is essential to consider that the use of Blockchain also brings with it a series of threats to cover, such as majority attacks or block validation attacks \cite{saad2020exploring}.}

\section{Conclusions and Future Work}
\label{sec:conclusion}


This work proposes a privacy-preserving framework for IoT malware detection that leverages FL to train and evaluate both supervised and unsupervised models without sharing sensitive data. This framework is designed to be deployed on the network nodes providing access to the IoT devices in Wifi, 5G or B5G networks, offloading the computation from the IoT device itself. \addtxtRev{In this sense, the client side is designed to be deployed on the RAN while the server side is intended for Fog/Cloud deployment.} To demonstrate its feasibility in a realistic IoT scenario, the N-BaIoT dataset has been used due to its heterogeneity and divisibility in terms of IoT devices and malware samples. Using N-BaIoT, we compared the performance of: i) a federated approach, where all device owners train their own model, which are periodically aggregated in a server, ii) a non privacy-preserving setup, in which the whole dataset is centralized and trained by the server, and iii) a local setup where each device owner trains one isolated and individual model. This comparison has shown that the use of more diverse and larger data, as done in the federated and centralized methods, has a considerable positive impact on the model performance both in a supervised and in an unsupervised scenario. 
Besides, it has been demonstrated that the privacy of the data can be preserved without losing model performance by following the federated approach. The resilience of the federated models against malicious clients has been tested through the following adversarial attacks: i) a data poisoning attack flipping all labels, ii) a model poisoning attack multiplying gradients by a negative factor, and iii) a model cancelling attack. The results showed that without using a robust technique to aggregate the models, a single malicious client in the federation can ruin the model.
Several robust aggregation functions, acting as countermeasures against adversarial attacks, have been applied to solve this problem, with median aggregation showing promising yet insufficient improvements. 
This first step in the direction of making the system robust against attacks shows that a lot of effort is still required to reach satisfying outcomes.

As future work, we plan to evaluate the impact of adversarial attacks in the unsupervised scenario to verify that they affect the results in a similar way as in the supervised counterpart. \addtxtRev{Moreover, testing the robustness of the model against evasion attacks, using forged adversarial samples to avoid detection at evaluation time, could also be an interesting future direction.}
Additionally, this work plans further research on the \addtxt{existing} countermeasures against adversarial attacks\addtxt{, such as Krum, Bulyan and AUROR.}

\addtxtRev{Scalability in real B5G scenarios is also a matter that could not be studied with any of the available datasets, raising a need for generating a much larger and much more diverse one.} \changeRev{Finally, t}{T}he deployment of the architecture in a fully distributed manner using Blockchain for the exchange of the federated models is also considered. \addtxtRev{Besides, Blockchain incorporation into the framework could improve possible security and privacy concerns of the clients.}



\section*{Acknowledgements}

This work has been partially supported by \textit{(a)} the Swiss Federal Office for Defense Procurement (armasuisse) with the TREASURE (R-3210/047-31) and CyberSpec (CYD-C-2020003) projects, by \textit{(b)} the European Commission through 5GZORRO project (Grant No. 871533) part of the 5G PPP in Horizon 2020, and by \textit{(c)} the University of Zürich UZH. 
We also thank Freepik for the icons used to represent the IoT devices in \figurename~\ref{fig:system_architecture}.

\bibliographystyle{IEEEtran}
\bibliography{references}


\end{document}